# Customized optical chirality of vortex structured light through state and degree of polarization control

Kayn A. Forbes* and Dale Green

*School of Chemistry, University of East Anglia, Norwich, Norfolk, NR7 4TJ, U.K.*

k.forbes@uea.ac.uk

**Abstract** We show how both the ellipticity $\eta$ and degree of polarization $P$ influences the extraordinary optical chirality properties of non-paraxial vortex beams. We find that, in stark contrast to paraxial optics and non-vortex modes, extremely rich and tuneable spatial distributions of optical chirality density can be produced by an optical vortex beam under tight focussing. We develop a theoretical description of how the optical chirality can be tailored for purpose by altering both the state $\eta$ and degree of polarization $P$ of the input vortex mode, along with the magnitude and sign of optical orbital angular momentum via the pseudoscalar topological charge $\ell$. We expect that the results will have a significant role in both producing novel techniques and improving existing methods in chiral nano-optics and structured light photonics.

**1. Introduction** Chiral objects are non-superimposable with their mirror image. Material chirality pervades the universe at all scales, from the spiralling arms of galaxies to our hands and feet, to the biomolecules responsible for life on earth. Light may also be chiral, most commonly manifest in circular polarization states, where the electromagnetic field vectors trace out helices that may twist to the left or right. This chirality associated with the handedness of polarization is often described in terms of the pseudoscalar optical helicity/chirality $\sigma = \pm 1$ where upper (lower) sign refers to left-handed (right-handed). A prevalent method of studying material chirality is to use chiral light in chiroptical spectroscopy, classic examples include circular dichroism, optical rotation, and vibrational optical activities [1,2]. The basis of such spectroscopy is extremely simple: a right-handed (left-handed) molecule or nanostructure interacts with a left-handed circularly polarized photon $\sigma = 1$ in a different way than it does with a right-handed circularly polarized photon $\sigma = -1$, this discrimination is referred to as natural optical activity [1]. The importance of chiroptical spectroscopy cannot be overstated: essentially all the molecular building blocks of life (nucleosides, amino acids, proteins, carbohydrates, etc.) are chiral; over half of developed pharmaceutical drugs are chiral [3]; and considering the recent challenge faced globally by COVID-19 it is pertinent to state that chiral spectroscopy can be used to study viruses [4,5]. It is therefore easy to appreciate why chiroptical spectroscopy is a widespread and flourishing area of research throughout science, being applied in chemical systems [1,5,6], biomolecules [7–10], and artificial nanostructures and metamaterials [11–16]. All these studies, while concerning an extremely diverse range of chiral materials, still predominantly rely on circularly polarized light as the chiral optical probe.

Structured light is a term which refers to the fact due to significant advances in optics technology we can tailor beams of light to possess inhomogeneous polarization, amplitude, and phase, both spatially and temporally [8,9]. One of the most well-known types of structured light is the optical vortex, a generic term which denotes an electromagnetic field that possesses an azimuthal phase $\exp(i\ell\phi)$, where $\ell \in \mathbb{Z}$ is the pseudoscalar topological charge and $\phi$ is the azimuthal coordinate. Specific modes include Laguerre-Gaussian and Bessel beams. Optical vortices have found immense application in a diverse range of areas, see latest reviews [17,19–24]. Optical vortices are chiral irrespective of their polarization: their helical wavefront (surface of constant phase) is chiral and can twist to the left $\ell > 0$ or right $\ell < 0$. Since 2018 there has been a surge in research activity applying the chirality of optical vortices in chiroptical light-matter interactions and spectroscopies [22,25]. Cutting-edge experiments include x-ray vortex dichroism of organometallic compounds [26], nonlinear vortex dichroism in chiral molecules as small as fenchone and limonene [27]; and Raman optical activity of vortex beams in liquid crystals [28].

Critical to engaging the chirality associated with the phase of an optical vortex is the consideration that it is a global property of the beam (its structure spans the transverse dimension of the beam, i.e. beam width), unlike polarization which is a local property. As such, in order for the chirality of materials to discriminate the handedness of the optical vortex, $\ell > 0$ or $\ell < 0$, the size-scale must be relatively similar: for small chiral nanostructures and molecules the light must therefore be spatially confined, e.g. by tight focusing. It is well known that the state and degree of polarization of the input beam significantly influences the electromagnetic fields around the focal plane of a non-paraxial field [29,30]. All such previous studies looking at the optical chirality density of vortex beams have been concerned with beams that are fully polarized (i.e. degree of polarization $P$ of the input beam before focussing is $P=1$) and of the two extremes of ellipticity $\eta$, i.e. linearly polarized $\eta = 0$ or circularly polarized $\eta = \pm \pi/4$. In this study we look specifically at how the degree of polarization and ellipticity influences the optical chirality density of vortex beams in nano-optics. It is highlighted that varying these two experimentally controllable parameters produces an extremely rich and diverse number of landscapes of optical chirality density in vortex beams which has hitherto gone unexploited.

**2. Optical Chirality Density of a non-paraxial Bessel beam**

**2.1 Electromagnetic fields of non-paraxial Bessel beams and optical chirality density** We use analytical theory to describe non-paraxial Bessel beams. We use the standard language first introduced by Lax et al. [31]. The zeroth-order transverse electromagnetic fields $\mathrm{T}_0$ for a Bessel beam propagating along $z$ are

$$\boldsymbol{E}^{(\mathrm{T}_0)} = (\alpha \hat{\boldsymbol{x}} + \beta \hat{\boldsymbol{y}}) J_\ell [k_t r] E_0 \, \mathrm{e}^{i(k_z z + \ell \phi - \omega t)}, \qquad \boldsymbol{B}^{(\mathrm{T}_0)} = (\alpha \hat{\boldsymbol{y}} - \beta \hat{\boldsymbol{x}}) \frac{k_z}{k} J_\ell [k_t r] B_0 \, \mathrm{e}^{i(k_z z + \ell \phi - \omega t)}, \qquad (1)$$

where $(\alpha, \beta)$ are the generalized Jones vectors, $E_0$ is the electric field amplitude, $B_0 = E_0/c$, $k^2 = \omega^2/c^2 = k_x^2 + k_y^2 + k_z^2 = k_t^2 + k_z^2$ is the square of the wave number, with $k_z = \sqrt{k^2 - k_t^2}$, $k_t = \sqrt{k_x^2 + k_y^2}$; $J_\ell [k_t r]$ are Bessel functions of the first kind; $\exp(i\ell\phi)$ is the azimuthal phase mentioned in the Introduction; and $\omega$ is the mean circular frequency. We drop dependences [] from now on for notational brevity. Both fields in (1) describe a paraxial mode and the electromagnetic fields of a Bessel beam correctly when $k \approx k_z$, e.g. a well-collimated beam. However, it is easy to show that (1) do not satisfy Maxwell's equations in their current form, e.g. $\nabla \cdot \boldsymbol{E}^{(\mathrm{T}_0)} \neq 0$. By using Maxwell's equations in a well-known method first developed by Lax et al. [31], we can generate the electromagnetic fields of a Bessel beam up to second order in a smallness parameter $(k_t/k_z)$, and thus they now include the zeroth-order transverse field $\mathrm{T}_0$; the first-order longitudinal field $\mathrm{L}_1$ (polarized along $\hat{z}$); and the second-order transverse fields $\mathrm{T}_2$ (see Supplementary Material for derivation):

$$\boldsymbol{E} = \begin{bmatrix} (\alpha\hat{\boldsymbol{x}}+\beta\hat{\boldsymbol{y}})J_\ell + \hat{\boldsymbol{z}}\dfrac{ik_t}{2k_z}\left(\{\alpha+i\beta\}e^{-i\phi}J_{\ell-1} + \{i\beta-\alpha\}e^{i\phi}J_{\ell+1}\right) \\ +\dfrac{k_t^2}{4k^2}\left(\hat{\boldsymbol{x}}\left[2\alpha J_\ell + J_{\ell-2}\{\alpha+i\beta\}e^{-2i\phi} + J_{\ell+2}\{\alpha-i\beta\}e^{2i\phi}\right]\right. \\ \left.+\hat{\boldsymbol{y}}\left[2\beta J_\ell + J_{\ell-2}\{i\alpha-\beta\}e^{-2i\phi} + J_{\ell+2}\{-\beta-i\alpha\}e^{2i\phi}\right]\right) \end{bmatrix} E_0 e^{i(k_z z + \ell\phi - \omega t)}, \qquad (2)$$

$$\boldsymbol{B} = \begin{bmatrix} (\alpha\hat{\boldsymbol{y}}-\beta\hat{\boldsymbol{x}})\dfrac{k_z}{k}J_\ell + \hat{\boldsymbol{z}}\dfrac{ik_t}{2k}\left(\{i\alpha-\beta\}e^{-i\phi}J_{\ell-1} + \{i\alpha+\beta\}e^{i\phi}J_{\ell+1}\right) \\ +\dfrac{k_t^2}{4kk_z}\left(\hat{\boldsymbol{x}}\left[-2\beta J_\ell + J_{\ell-2}\{i\alpha-\beta\}e^{-2i\phi} + J_{\ell+2}\{-i\alpha-\beta\}e^{2i\phi}\right]\right. \\ \left.+\hat{\boldsymbol{y}}\left[2\alpha J_\ell + J_{\ell-2}\{-i\beta-\alpha\}e^{-2i\phi} + J_{\ell+2}\{i\beta-\alpha\}e^{2i\phi}\right]\right) \end{bmatrix} B_0 e^{i(k_z z + \ell\phi - \omega t)}. \qquad (3)$$

Both (2) and (3) can accurately describe a paraxial or non-paraxial (e.g. tightly focused) Bessel beam. The smallness parameter for a Bessel beam is $k_t/k_z$: increasing the size of this factor essentially accounts for tightly focusing the beam. For example, in the far field $k_t/k_z \approx 0$ (or $k \approx k_z$) for a z-propagating beam and the field can essentially be described by the zeroth-order transverse fields in (2) and (3), or equally (1) to an almost exact approximation (i.e. a paraxial description) [32]. By focusing the beam $k_t/k_z$ becomes larger and the additional longitudinal and transverse terms in (2) and (3) become important in magnitude and are responsible for the extraordinary properties of structured light in nano-optics [33–35]. Thus, the paraxial description (1) no long suffices under tight-focussing. Throughout this paper we refer to the polarization properties of the input light in the far field as two-dimensional (2D), i.e., the zeroth-order electric field polarization state. The polarization properties of the non-paraxial field under spatial confinement are generally referred to as three-dimensional (3D) polarization [36–38].

The optical chirality density $C$ for a quasi-monochromatic beam may be defined as (see Supplementary Material for further information) [39–41]

$$C = -\dfrac{\varepsilon_0 \omega}{2}\operatorname{Im}(\bar{\boldsymbol{E}}\cdot\boldsymbol{B}), \qquad (4)$$

where the overbar denotes complex conjugation. Crudely put, this dynamic property of light (conserved in free space) gives a measure of how chiral the optical field is. It is related to the optical helicity, interested readers are referred to refs [39,41] for further information. Chiral light-matter interactions are produced from multipolar interferences that give space-odd, time-even tensors for the material [1,42]. It is of utmost importance to appreciate (4) couples to the electric dipole magnetic dipole (E1M1) interferences of chiral materials. The optical chirality, defined by (4) therefore does not account for all chiral light-matter interactions, for example those which stem from electric dipole electric quadrupole interferences (E1E2). Inserting (2) and (3) into (4) gives the optical chirality density as (full systematic derivation in SI)

$$C = -\frac{I\omega}{c^2}\left[\frac{k_z}{k}J_\ell^2 P\sin 2\eta + \frac{k_t^2}{4kk_z}\left(2\left\{1+\frac{k_z^2}{k^2}\right\}J_\ell^2 P\sin 2\eta + \{P\sin 2\eta +1\}J_{\ell-1}^2 + \{P\sin 2\eta -1\}J_{\ell+1}^2\right)\right.$$

$$\left. +\left(\frac{k_t^4}{8k^3 k_z}\right)\left(2J_\ell^2 P\sin 2\eta + J_{\ell-2}^2\{P\sin 2\eta +1\} + J_{\ell+2}^2\{P\sin 2\eta -1\}\right)\right], \quad (5)$$

where $I = c\varepsilon_0 E_0^2/2$ is the intensity of the beam, $P$ is the 2D degree of polarization of the input, and $\eta$ defines the degree of 2D polarization ellipticity: $\eta = 0$ is linearly polarized light, $\eta = \pm\pi/4$ is pure circular, and $-\pi/4 < \eta < \pi/4$ corresponds to elliptically polarized light (positive signed is right-handed, negative sign left-handed). The first term in square brackets in (5) corresponds to the well-known optical chirality density for a paraxial beam of light which stems purely from a degree of ellipticity in the 2D polarization state, proportional to the third Stokes parameter, and clearly requires $P \neq 0$. All the other terms correspond to the optical chirality density generated by first-order longitudinal and second-order transverse fields which become important under the non-paraxial conditions we are interested in. Note that no terms in (5) depend on the azimuth $\theta$, i.e. the orientation of the 2D polarization ellipse.

**2.2 Optical chirality density of a Bessel beam with pure $P$=1 2D Polarization** The optical chirality density (5) is plotted in Figures 1 and 2 for $P = 1$ and $k_t/k_z = 0.6315$ (i.e. tightly focused).

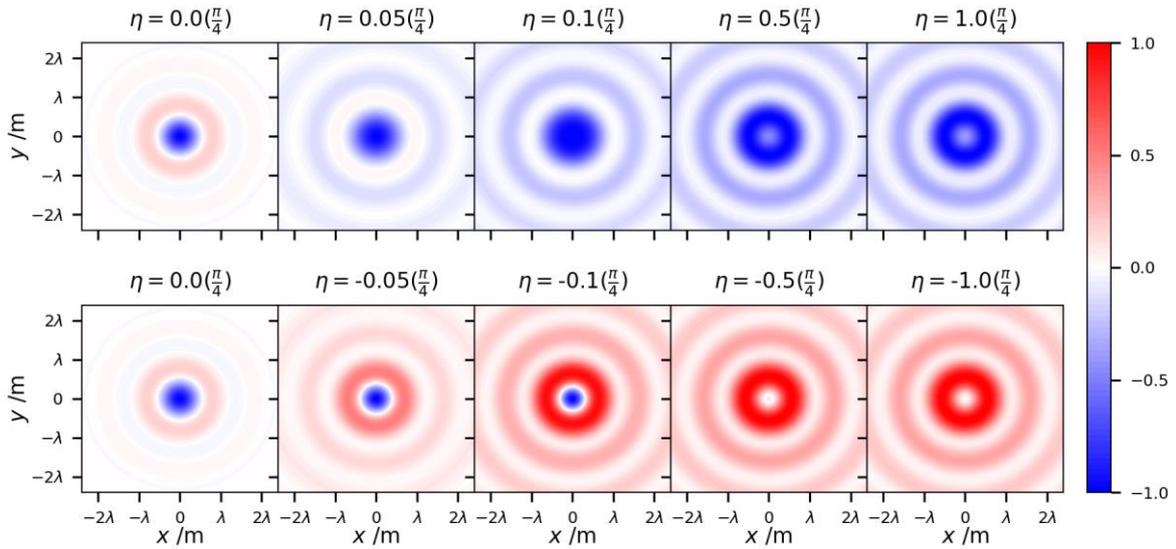

**Figure 1:** Evolution of the optical chirality density of an $\ell = 1$ Bessel beam in the focal plane with varying 2D-polarization ellipticity $\eta$. The input polarization progresses from linearly polarized $\eta = 0$ through varying degrees of ellipticity $-\pi/4 < \eta < \pi/4$ until it reaches pure circularly polarization $\eta = \pm\pi/4$. The top row cycles through right-handed polarization; the bottom row left-handed polarization. In all plots $k_t/k_z = 0.6315$ and each plot is normalized individually.

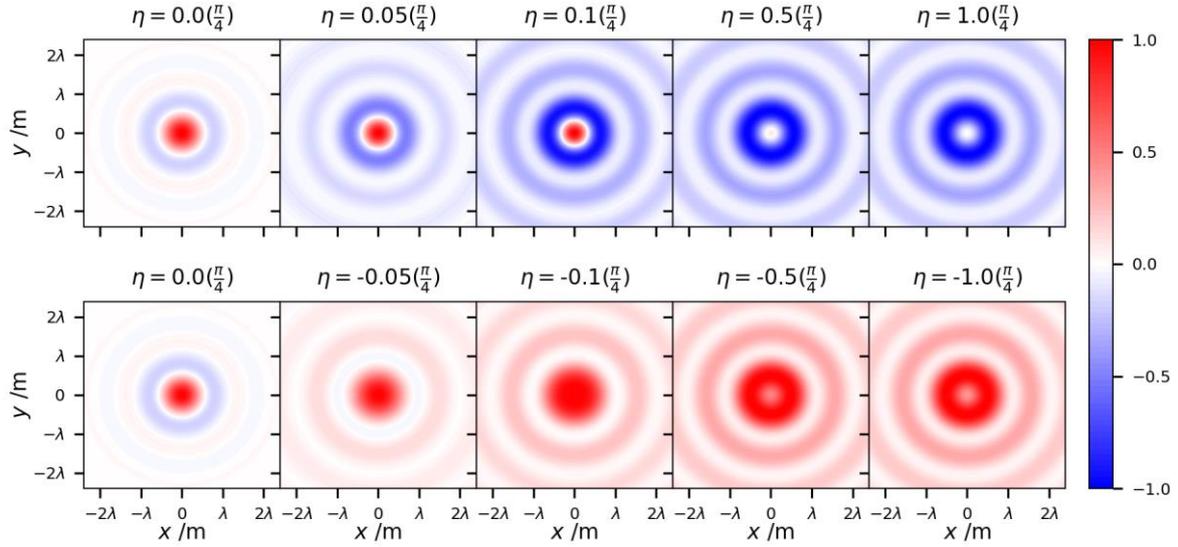

**Figure 2:** Evolution of the optical chirality density of an $\ell = -1$ Bessel beam in the focal plane with varying 2D-polarization ellipticity $\eta$. (Everything else same as Figure 1).

The leftmost column in Figures 1 and 2 corresponds to an input 2D linearly polarized ($\eta = 0$) Bessel beam, the rightmost column corresponds to a circularly polarized Bessel beam, and in between shows varying degrees of ellipticity. Firstly, we note that 2D linearly polarized vortex beams under tight focussing exhibit non-zero optical chirality densities [44,45]: this is in stark contrast to plane waves or paraxial beams, where there must be a degree of ellipticity $\eta$ in the polarization state. Furthermore, comparing Figure 1 and 2 shows that the optical chirality density spatial distributions reverses depending on the sign of $\ell$, i.e. the vortex wavefront handedness. It is this non-zero chirality associated with tightly focused linearly polarized vortex beams which is responsible for vortex dichroism [26,46] and vortex differential scattering [47]: chiral materials absorb and scatter linearly polarized non-paraxial vortex beams at different rates depending on whether the input beam is $\ell > 0$ or $\ell < 0$.

Moving across the rows in Figures 1 and 2 we are changing the state of 2D polarization by increasing the degree of 2D ellipticity of the input and this significantly alters the spatial distribution of optical chirality density. It is noticeable that we soon lose half of the rings of chirality density (the left-hand panels have 5 rings for example, whereas the rightmost have 3). We also note that there is a chiral interplay between the signs of $\ell$ (vortex handedness) and the sign of $\eta$ (polarization handedness): this is due to spin-orbit interactions of light. Most striking is the fact that when the handedness of the vortex and polarization are opposite (e.g. left-handed vortex, right-handed ellipticity) we produce on-axis optical chirality densities of the same sign; where they are of the same handedness there is an opposite signed chirality density which progresses to a null density along the axis. This is somewhat similar to the behaviour of the intensity of tightly focused vortex beams, where it is often described in terms of parallel and antiparallel spin and orbital angular momentum of the beams [34,48]. Furthermore, in the cases of $\eta \approx \pm\pi/40, \ell \pm 1$ (i.e. we have antiparallel polarization and vortex handedness) we produce the spatial distribution of the optical chirality of an $\ell = 0$ mode (see Figure 3 for example), even though $\ell \neq 0$.

It is simple to highlight the significant influence optical OAM and vortex chirality has on optical chirality by plotting (5) for $\ell = 0$, i.e. a tightly focussed non-vortex beam with no OAM. Figure 3 highlights that when there is no degree of ellipticity in the 2D polarization state ($\eta = 0$) the optical chirality is zero, and while the magnitude of the optical chirality density increases with increasing ellipticity, the spatial distributions are invariant to both magnitude and sign (handedness). This is clearly in stark contrast to tightly focussed $\ell \neq 0$ Bessel modes which possess OAM in Figs 1 and 2.

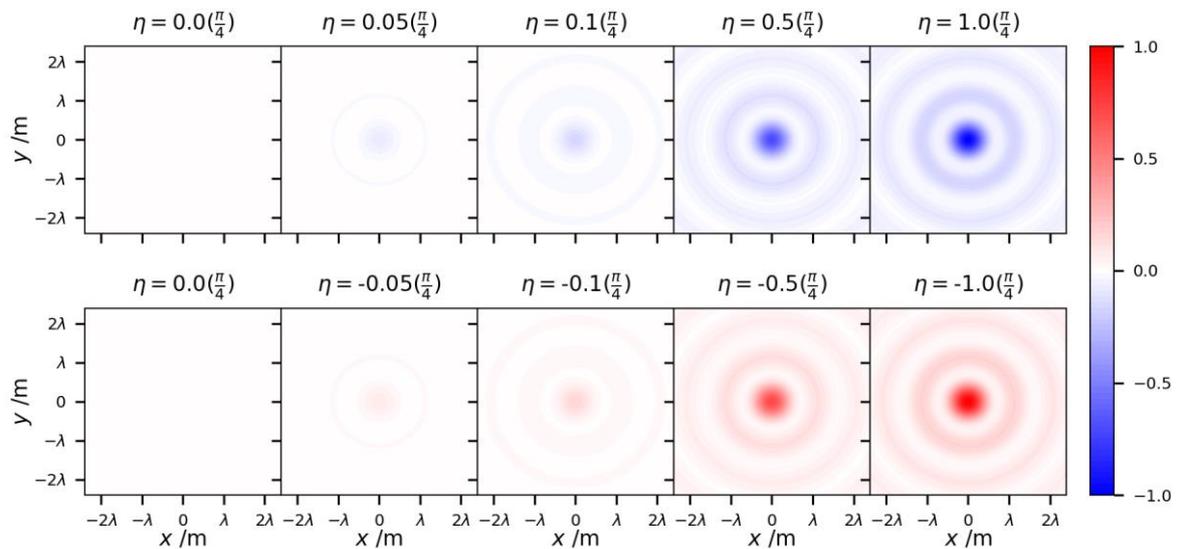

**Figure 3:** Evolution of the optical chirality density of a tightly focussed $\ell = 0$ Bessel beam in the focal plane with varying 2D-polarization ellipticity $\eta$. Each plot is normalized against $|\eta| = \pi/4$ plots. (Everything else same as Figure 1).

Another degree of freedom we can control is the magnitude of $\ell$ for the input beam, and plots for $\ell = \pm 2$ modes can be found in the Supplementary Material. The optical chirality spatial distributions follow a similar pattern to Figs 1 and 2 for the larger value of $\ell$, the main difference being the increase in ring widths analogous to the behaviour of the intensity of vortex beams.

In order to appreciate the role tight focusing has in producing the extraordinary optical chirality properties highlighted in Figures 1 and 2 it is useful to plot (5) under paraxial conditions. We gave a qualitative physical reason for the necessity for tight focusing in the Introduction: here we see it played out by the mathematics. The analogous plots of Figure 1 and Figure 2 for a weakly focussed $(k_t / k_z) = 0.01$, essentially paraxial beam, can be found in Figure 4, where it is readily observed all the extraordinary properties of the non-paraxial field are lost. Indeed, there are no observable spin-orbit interactions; there is a vanishingly small (practically zero) chirality for linearly polarised inputs $\eta = 0$; and the optical chirality density does not depend on the sign of vortex handedness: the plots in Figure 4 are identical for $\ell = \pm 1$.

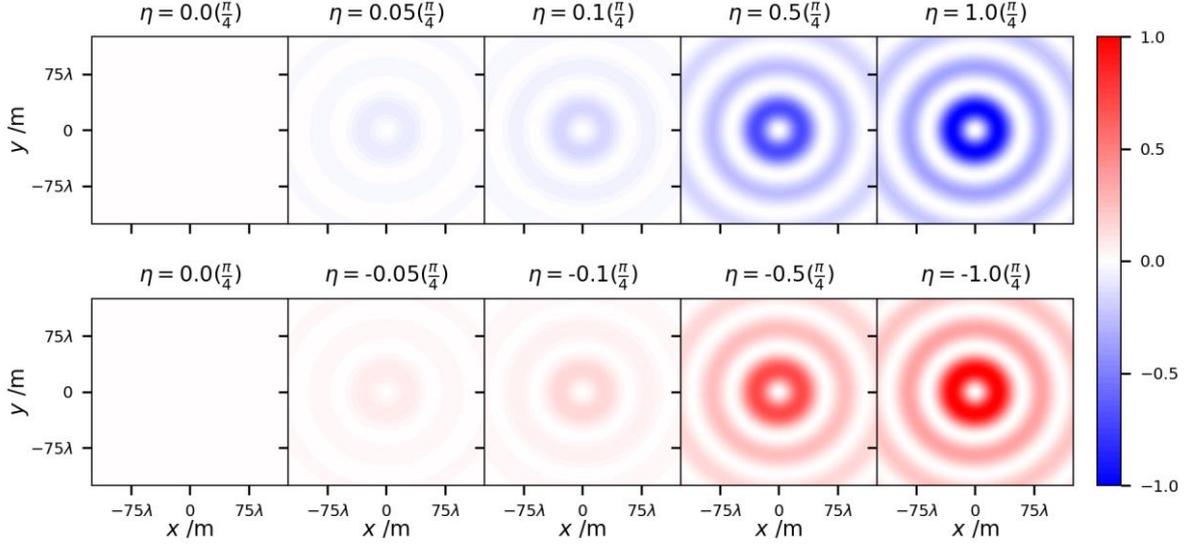

**Figure 4:** Evolution of the optical chirality density of an $\ell = 1$ and $\ell = -1$ weakly focussed $(k_t / k_z) = 0.01$ Bessel beam in the focal plane with varying 2D-polarization ellipticity $\eta$. Each plot is normalized against $|\eta| = \pi/4$ plots. (Everything else same as Figure 1).

**2.3 Optical chirality density of a Bessel beam with partial 2D polarization** Between the two extremes of completely polarized light $P = 1$ and unpolarized light $P = 0$ we have partially polarized light. In the previous section we studied how the state of polarization (specifically the degree of ellipticity $\eta$) affects the optical chirality density in the focal plane for a fully polarized beam $P = 1$; in this section we will study how the degree of polarization $P$ effects the optical chirality density for a set of fixed polarization states: Figure 5 $\eta = \pi/80$; Figure 6 $\eta = \pi/40$; and Figure 7 $\eta = \pi/8$.

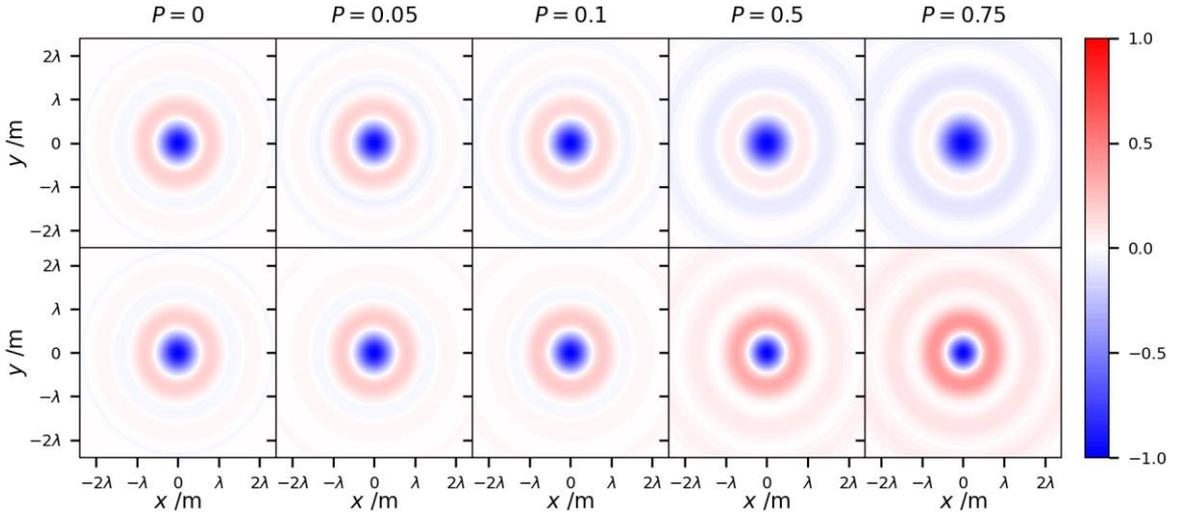

**Figure 5:** Evolution of the optical chirality density of an $\ell = 1$ Bessel beam in the focal plane with varying degree of polarization $P$ of an elliptically polarized beam $\eta = \pm\pi/80$. The top row cycles through right-handed polarization $\eta = \pi/80$; the bottom row left-handed polarization $\eta = -\pi/80$. In all plots $k_t / k_z = 0.6315$ and each plot is normalized individually.

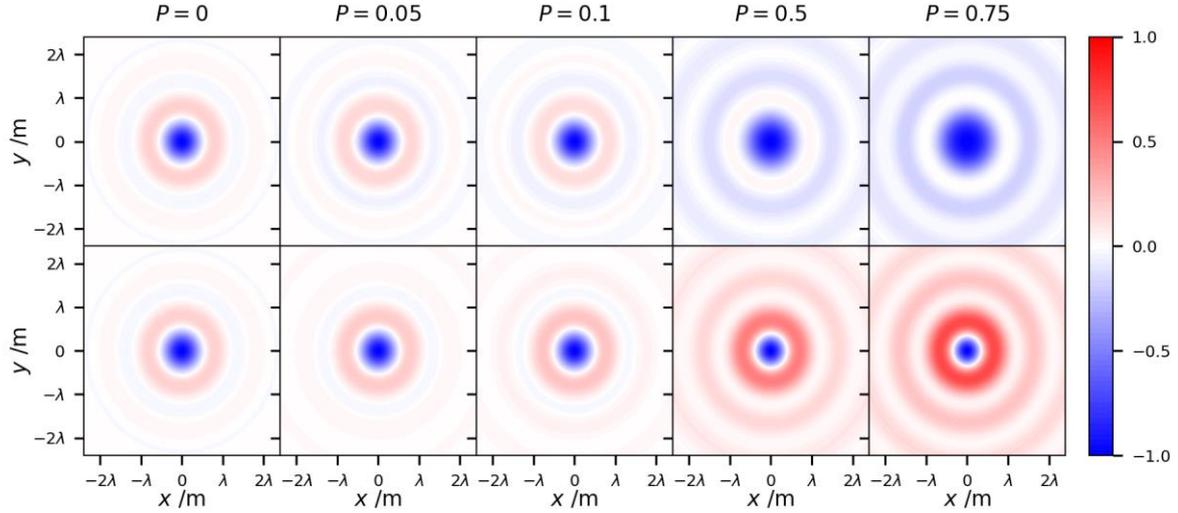

**Figure 6:** Evolution of the optical chirality density of an $\ell = 1$ Bessel beam in the focal plane with varying degree of polarization $P$ of an elliptically polarized beam $\eta = \pm\pi/40$. Everything else is as Figure 5.

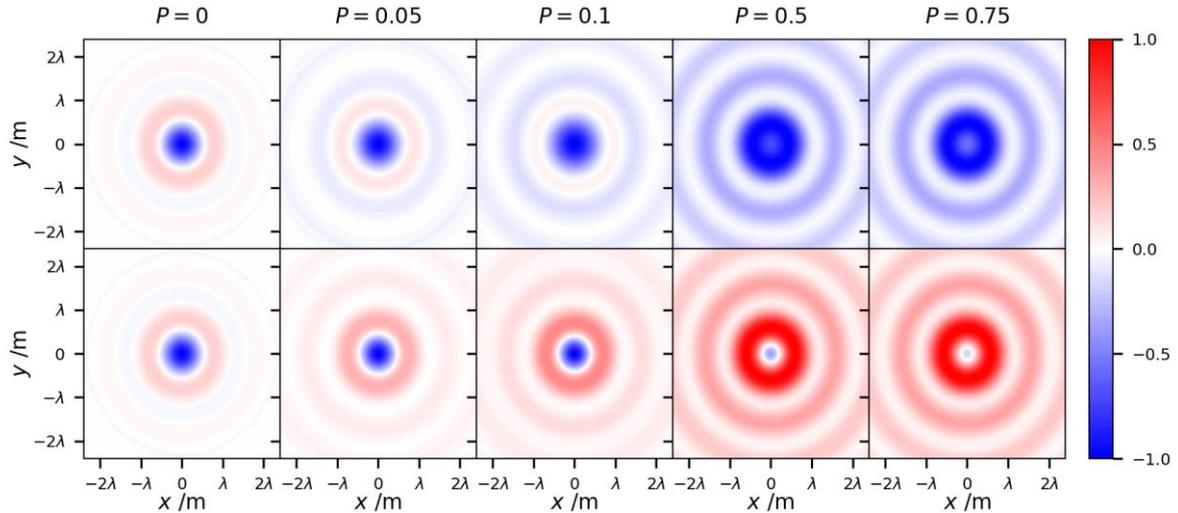

**Figure 7:** Evolution of the optical chirality density of an $\ell = 1$ Bessel beam in the focal plane with varying degree of polarization $P$ of an elliptically polarized beam $\eta = \pm\pi/8$. Everything else is as Figure 5.

First we must point out the remarkable result that even when the beam is unpolarized $P = 0$ (leftmost columns in Figs 5-7), the optical chirality density is non-zero [50]. In other words, we can get chiral light-matter interactions and do chiral spectroscopy with unpolarized sources of light [43,50,51]. Furthermore, Figures 5-7 clearly highlight how increasing (or decreasing) the degree of 2D polarization $P$ significantly influences the spatial distribution of the optical chirality density, even though the state of 2D polarization $\eta$ for the polarized part of the field is fixed in each case (and so too is the value of $\ell$).

Another way to gauge the influence of $P$ is to compare how the spatial distribution of the optical chirality of a fully 2D polarized beam for a given $\eta$ (from Figures 1 and 2) compares to that where $P \neq 1$ (Figures 5-7). As an example, Figure 7 shows the optical chirality density of a partially polarized beam with $\eta = \pi/8$, which can be compared to the fourth column in Figure 1. We see that Figure 7 shows that a $P = 0.1$, $\eta = \pi/8$ gives a similar optical chirality density as a $P = 1$, $\eta = \pi/80$ beam. In this section we have concentrated on $\ell=1$ modes, however the corresponding $\ell=-1$ plots can be found in the Supplementary Material, alongside the $\ell=\pm 2$ mode plots.

**3. Discussion and Conclusion** Here we have systematically studied how the 2D state and degree of polarization of an input paraxial vortex beam influences the optical chirality density of the non-paraxial field around the focal plane. We have highlighted the extremely rich spatial distributions of optical chirality density that can be produced by optical vortex beams in nano-optics. This optical chirality density can be tailored for purpose by altering both the state and degree of 2D polarization of the input vortex mode, along with the magnitude and sign of OAM through $\ell$. It must be remembered that the optical chirality density produced by paraxial beams or plane waves requires both $P \neq 0$ and $\eta \neq 0$, and altering these parameters does not lead to any observable variations in the spatial distributions of the optical chirality density, they just vary in magnitude (e.g. Figure 4). Comparison between this currently prevalent optical probe in chiroptical spectroscopy versus the optical chirality of non-paraxial vortex beams we have highlighted in this work makes it readily clear why structured light chirality is poised to revolutionize chiral light-matter interactions [22,25,52].

It is worth briefly discussing another extraordinary property of optical fields in nano-optics at this juncture: the transverse spin momentum of light [33,53]. There have been a number of recent studies looking at the influence of the state and degree of 2D polarization on this property [54–57]. Like the optical chirality density of optical vortex beams, the transverse spin momentum density of light occurs even if the optical source is 2D unpolarized [54,55]. This property of transverse spin is extremely generic for confined electromagnetic fields, being present in surface evanescent waves and tightly focused laser beams, for example. However, the optical chirality density which exists for 2D unpolarized light strictly only occurs in optical vortex beams [50]. It is rather interesting to reflect on the extraordinary fact that light can possess both spin angular momentum and optical chirality density for 2D unpolarized fields, and that the state and degree of 2D polarization significantly influences the spatial distributions of these quantities.

As previously mentioned in the Introduction, to date practically all chiroptical spectroscopy and chiral light-matter interactions utilize circularly polarized light (e.g., unstructured Gaussian beams) as the chiral optical probe. The optical chirality density for vortex beams we have studied has already proven to offer significant advancements in both generating novel and improving existing applications in chiral nano-optics [25–28,43]. Here we have shown that this extraordinary optical chirality of structured light is able to be even further tailored for purpose than hitherto thought. Future studies are aimed towards other forms of structured light which have enhanced chiral properties [58], such as vector vortex modes, alongside the influence of second-order coherence.

# Supplementary Material

# Customized optical chirality of vortex structured light through state and degree of polarization control

Kayn A. Forbes* and Dale Green

*School of Chemistry, University of East Anglia, Norwich, Norfolk, NR7 4TJ, U.K.*

k.forbes@uea.ac.uk

**Derivation of Bessel beam electromagnetic fields up to second order in smallness parameter**

The derivation of Equations (2) and (3) from the main manuscript are as follows. The zeroth order transverse $T_0$ electromagnetic fields for a Bessel beam are well known:

$$\boldsymbol{E}^{(T_0)}[\boldsymbol{r},t] = (\alpha\hat{\boldsymbol{x}} + \beta\hat{\boldsymbol{y}}) J_\ell[k_t r] E_0 \, e^{i(k_z z + \ell\phi - \omega t)} \tag{1.6}$$

$$\boldsymbol{B}^{(T_0)}[\boldsymbol{r},t] = (\alpha\hat{\boldsymbol{y}} - \beta\hat{\boldsymbol{x}}) \frac{k_z}{k} J_\ell[k_t r] B_0 \, e^{i(k_z z + \ell\phi - \omega t)} \tag{1.7}$$

We assume quasi-monochromatic frequencies centred upon the average frequency $\omega$. All definitions of symbols and notation are given in the main manuscript. Neither of these fields satisfy Maxwell's equations. This is simple enough to see by inserting them into Gauss's Law $\nabla \cdot \boldsymbol{E}^{(T_0)} \neq 0$ and $\nabla \cdot \boldsymbol{B}^{(T_0)} \neq 0$. To remedy this one can use the method first developed by Melvin Lax et al.[1]

$$\begin{aligned}\nabla \cdot \boldsymbol{E}^{(T_0)} &= \left(\hat{\boldsymbol{x}}\frac{\partial}{\partial x} + \hat{\boldsymbol{y}}\frac{\partial}{\partial y} + \hat{\boldsymbol{z}}\frac{\partial}{\partial z}\right) \cdot \boldsymbol{E}^{(T_0)} \\ &= \nabla_\perp \cdot \boldsymbol{E}^{(T_0)} + \hat{\boldsymbol{z}}\frac{\partial}{\partial z} \cdot \boldsymbol{E}^{(T_0)} \\ &= 0.\end{aligned} \tag{1.8}$$

We can re-arrange (1.8) to calculate the longitudinal component $E_z$ via

$$E_z = -\int \nabla_\perp \cdot \boldsymbol{E}^{(T_0)} \partial z. \tag{1.9}$$

The same manipulation for the magnetic field gives

$$B_z = -\int \nabla_\perp \cdot \boldsymbol{B}^{(T_0)} \partial z. \tag{1.10}$$

Now we need to insert (1.6) and (1.7) into (1.9) and (1.10), respectively to yield

$$E_z = \frac{i}{k}\left(\alpha\frac{\partial}{\partial x} + \beta\frac{\partial}{\partial y}\right) J_\ell[k_t r] E_0 \, e^{i(k_z z + \ell\phi - \omega t)}, \tag{1.11}$$

and

$$B_z = \frac{i}{k}\left(\alpha \frac{\partial}{\partial y} - \beta \frac{\partial}{\partial y}\right)\frac{k_z}{k} J_\ell[k_t r] B_0 \, e^{i(k_z z + \ell \phi - \omega t)}.$$

(1.12)

From now on we concentrate on the electric field derivation in the knowledge that the magnetic field follows analogous algebraic manipulations. We use the following Cartesian to cylindrical coordinate transformations throughout the derivations: $\frac{\partial}{\partial x} = (\cos\phi)\frac{\partial}{\partial r} - \frac{1}{r}(\sin\phi)\frac{\partial}{\partial \phi}$ and $\frac{\partial}{\partial y} = (\sin\phi)\frac{\partial}{\partial r} + \frac{1}{r}(\cos\phi)\frac{\partial}{\partial \phi}$.

$$\begin{aligned}
E_z &= \frac{i}{k_z}\left(\alpha\left\{(\cos\phi)\frac{\partial}{\partial r} - \frac{1}{r}(\sin\phi)\frac{\partial}{\partial \phi}\right\} + \beta\left\{(\sin\phi)\frac{\partial}{\partial r} + \frac{1}{r}(\cos\phi)\frac{\partial}{\partial \phi}\right\}\right) J_\ell[k_t r] E_0 \, e^{i(k_z z + \ell\phi - \omega t)} \\
&= \frac{i}{k_z}\left(\alpha\left\{(\cos\phi)\left\{\frac{k_t}{2}\left[J_{\ell-1}[k_t r] - J_{\ell+1}[k_t r]\right]\right\} - \frac{i\ell}{r} J_\ell[k_t r]\sin\phi\right\}\right. \\
&\quad \left. + \beta\left\{(\sin\phi)\left\{\frac{k_t}{2}\left[J_{\ell-1}[k_t r] - J_{\ell+1}[k_t r]\right]\right\} + \frac{i\ell}{r} J_\ell[k_t r]\cos\phi\right\}\right) E_0 \, e^{i(k_z z + \ell\phi - \omega t)} \\
&= \frac{i}{k_z}\left(\alpha\left\{(\cos\phi)\left\{\frac{k_t}{2}\left[J_{\ell-1}[k_t r] - J_{\ell+1}[k_t r]\right]\right\} - i\left\{\frac{k_t}{2}\left[J_{\ell-1}[k_t r] + J_{\ell+1}[k_t r]\right]\right\}\sin\phi\right\}\right. \\
&\quad \left. + \beta\left\{(\sin\phi)\left\{\frac{k_t}{2}\left[J_{\ell-1}[k_t r] - J_{\ell+1}[k_t r]\right]\right\} + i\left\{\frac{k_t}{2}\left[J_{\ell-1}[k_t r] + J_{\ell+1}[k_t r]\right]\right\}\cos\phi\right\}\right) E_0 \, e^{i(k_z z + \ell\phi - \omega t)} \\
&= \frac{ik_t}{2k_z}\left(\alpha\left\{(\cos\phi)\left\{\left[J_{\ell-1}[k_t r] - J_{\ell+1}[k_t r]\right]\right\} - i\left\{\left[J_{\ell-1}[k_t r] + J_{\ell+1}[k_t r]\right]\right\}\sin\phi\right\}\right. \\
&\quad \left. + \beta\left\{(\sin\phi)\left\{\left[J_{\ell-1}[k_t r] - J_{\ell+1}[k_t r]\right]\right\} + i\left\{\left[J_{\ell-1}[k_t r] + J_{\ell+1}[k_t r]\right]\right\}\cos\phi\right\}\right) E_0 \, e^{i(k_z z + \ell\phi - \omega t)} \\
&= \frac{ik_t}{2k_z}\left(\alpha J_{\ell-1}[k_t r]\cos\phi - \alpha i J_{\ell-1}[k_t r]\sin\phi - \alpha J_{\ell+1}[k_t r]\cos\phi - \alpha i J_{\ell+1}[k_t r]\sin\phi\right. \\
&\quad \left. + \beta J_{\ell-1}[k_t r]\sin\phi + i\beta J_{\ell-1}[k_t r]\cos\phi - \beta J_{\ell+1}[k_t r]\sin\phi + i\beta J_{\ell+1}[k_t r]\cos\phi\right) E_0 \, e^{i(k_z z + \ell\phi - \omega t)} \\
&= \frac{ik_t}{2k_z}\left[(\alpha + i\beta) J_{\ell+1}[k_t r] e^{-i\phi} + (i\beta - \alpha) J_{\ell-1}[k_t r] e^{i\phi}\right] E_0 \, e^{i(k_z z + \ell\phi - \omega t)}.
\end{aligned}$$

(1.13)

We have used the relation $J'_v(x) = \frac{1}{2}[J_{v-1}(x) - J_{v+1}(x)]$ which with our Bessel function's argument is $\frac{\partial}{\partial r} J(k_t r) = \frac{k_t}{2}[J_{v-1}(k_t r) - J_{v+1}(k_t r)]$ (making use of the chain rule) in the first manipulation; and we also use the Bessel relation $\frac{2v}{x} J_v(x) = J_{v+1}(x) + J_{v-1}(x)$ which specifically for us means $\ell J_v(x) = \frac{k_t r}{2}(J_{v+1}(k_t r) + J_{v-1}(k_t r))$ in the second manipulation, and finally Euler's formula at the end.

Inserting (1.13) into (1.8) gives the correct divergence free nature of the electric field as required by Maxwell's equations:

$$\boldsymbol{E}^{(T_0)} + \boldsymbol{E}^{(L_0)} = \left[ \underbrace{(\alpha\hat{\boldsymbol{x}} + \beta\hat{\boldsymbol{y}})J_\ell}_{\text{Zeroth-order Transverse } T_0} + \underbrace{\hat{\boldsymbol{z}}\frac{ik_t}{2k_z}\left(\{\alpha + i\beta\}e^{-i\phi} J_{\ell-1} + \{i\beta - \alpha\}e^{i\phi} J_{\ell+1}\right)}_{\text{First-order Longitudinal } L_0} \right] E_0\, e^{i(k_z z + \ell\phi - \omega t)}.$$

(1.14)

The magnetic field calculation follows a very similar derivation:

$$\boldsymbol{B}^{(T_0)} + \boldsymbol{B}^{(L_1)} = \left[ \underbrace{(\alpha\hat{\boldsymbol{y}} - \beta\hat{\boldsymbol{x}})\frac{k_z}{k}J_\ell}_{\text{Zeroth-order Transverse } T_0} + \underbrace{\hat{\boldsymbol{z}}\frac{ik_t}{2k}\left(\{i\alpha - \beta\}e^{-i\phi} J_{\ell-1} + \{i\alpha + \beta\}e^{i\phi} J_{\ell+1}\right)}_{\text{First-order Longitudinal } L1} \right] B_0\, e^{i(k_z z + \ell\phi - \omega t)} \quad (1.15)$$

Note we have dropped the argument of the Bessel functions now for notational brevity: we do this throughout the Supplementary Information from this point forward. To calculate the second order transverse electric fields $T_2$ we require the use of the Maxwell-Ampere Law:

$$\nabla \times \boldsymbol{B} = \frac{1}{c^2}\frac{\partial \boldsymbol{E}}{\partial t}. \quad (1.16)$$

To get the second order transverse electric field components $\boldsymbol{E}^{(T_2)}$ we insert the magnetic field (1.15) into the re-arranged Maxwell-Ampere law (1.16):

$$\boldsymbol{E} = c^2 \int (\nabla \times \boldsymbol{B})\partial t$$
$$= c^2 \frac{i}{\omega}\left(\nabla \times \left[(\alpha\hat{\boldsymbol{y}} - \beta\hat{\boldsymbol{x}})\frac{k_z}{k}J_\ell + \hat{\boldsymbol{z}}\frac{ik_t}{2k}\left(\{i\alpha - \beta\}e^{-i\phi} J_{\ell-1} + \{i\alpha + \beta\}e^{i\phi} J_{\ell+1}\right)\right] B_0\, e^{i(k_z z + \ell\phi - \omega t)}\right). \quad (1.17)$$

We generate the second order electric field components from the transverse gradient of the first order magnetic longitudinal field (the curl operator acting on the zeroth-order transverse magnetic field produces the zeroth-order transverse and first-order longitudinal electric field components which we already derived above using the other Maxwell equations).

$$\boldsymbol{E}^{(T_2)} = c^2 \frac{i}{\omega} \left( \left( \hat{\boldsymbol{x}} \frac{\partial}{\partial x} + \hat{\boldsymbol{y}} \frac{\partial}{\partial y} \right) \times \left[ \hat{\boldsymbol{z}} \frac{ik_t}{2k} \left( \{i\alpha - \beta\} e^{-i\phi} J_{\ell-1} + \{i\alpha + \beta\} e^{i\phi} J_{\ell+1} \right) \right] B_0 \, e^{i(k_z z + \ell \phi - \omega t)} \right)$$

$$= c^2 \frac{i}{\omega} \left( \left( \hat{\boldsymbol{x}} \frac{\partial}{\partial y} - \hat{\boldsymbol{y}} \frac{\partial}{\partial x} \right) \times \left[ \frac{ik_t}{2k} \left( \{i\alpha - \beta\} e^{-i\phi} J_{\ell-1} + \{i\alpha + \beta\} e^{i\phi} J_{\ell+1} \right) \right] B_0 \, e^{i(k_z z + \ell \phi - \omega t)} \right)$$

$$= c^2 \frac{i}{\omega} \left( \left( \hat{\boldsymbol{x}} \left[ (\sin\phi) \frac{\partial}{\partial r} + \frac{1}{r}(\cos\phi) \frac{\partial}{\partial \phi} \right] - \hat{\boldsymbol{y}} \left[ (\cos\phi) \frac{\partial}{\partial r} - \frac{1}{r}(\sin\phi) \frac{\partial}{\partial \phi} \right] \right) \right.$$
$$\left. \times \left[ \frac{ik_t}{2k} \left( \{i\alpha - \beta\} e^{-i\phi} J_{\ell-1} + \{i\alpha + \beta\} e^{i\phi} J_{\ell+1} \right) \right] \right) B_0 \, e^{i(k_z z + \ell \phi - \omega t)}$$

$$= c^2 \frac{i}{\omega} \left( \underbrace{\hat{\boldsymbol{x}} (\sin\phi) \frac{\partial}{\partial r} \left[ \frac{ik_t}{2k} \left( \{i\alpha - \beta\} e^{-i\phi} J_{\ell-1} + \{i\alpha + \beta\} e^{i\phi} J_{\ell+1} \right) \right]}_{1} \right.$$
$$+ \underbrace{\hat{\boldsymbol{x}} \frac{1}{r}(\cos\phi) \frac{\partial}{\partial \phi} \left[ \frac{ik_t}{2k} \left( \{i\alpha - \beta\} e^{-i\phi} J_{\ell-1} + \{i\alpha + \beta\} e^{i\phi} J_{\ell+1} \right) \right]}_{2}$$
$$\underbrace{- \hat{\boldsymbol{y}} (\cos\phi) \frac{\partial}{\partial r} \left[ \frac{ik_t}{2k} \left( \{i\alpha - \beta\} e^{-i\phi} J_{\ell-1} + \{i\alpha + \beta\} e^{i\phi} J_{\ell+1} \right) \right]}_{3}$$
$$\left. + \underbrace{\hat{\boldsymbol{y}} \frac{1}{r}(\sin\phi) \frac{\partial}{\partial \phi} \left[ \frac{ik_t}{2k} \left( \{i\alpha - \beta\} e^{-i\phi} J_{\ell-1} + \{i\alpha + \beta\} e^{i\phi} J_{\ell+1} \right) \right]}_{4} \right) B_0 \, e^{i(k_z z + \ell \phi - \omega t)}. \quad (1.18)$$

To keep track of calculations we split the algebraic manipulations up into the *x* and *y* components, 1, 2, 3, and 4. The first *x* component, labelled 1 in the final line of (1.18), is given by

$$\hat{\boldsymbol{x}} (\sin\phi) \frac{\partial}{\partial r} \left[ \frac{ik_t}{2k} \left( \{i\alpha - \beta\} e^{-i\phi} J_{\ell-1} + \{i\alpha + \beta\} e^{i\phi} J_{\ell+1} \right) \right] B_0 \, e^{i(k_z z + \ell \phi - \omega t)}$$
$$= \hat{\boldsymbol{x}} (\sin\phi) \left[ \frac{ik_t}{2k} \left( \{i\alpha - \beta\} e^{-i\phi} \left\{ \frac{k_t}{2}(J_{\ell-2} - J_\ell) \right\} + \{i\alpha + \beta\} e^{i\phi} \left\{ \frac{k_t}{2}(J_\ell - J_{\ell+2}) \right\} \right) \right] B_0 \, e^{i(k_z z + \ell \phi - \omega t)}, \quad (1.19)$$

where we used the Bessel relation $J'(x) = \frac{1}{2}\left[ J_{v-1}(x) - J_{v+1}(x) \right]$ in a similar way to earlier. The second *x* component is:

$$\hat{x}\frac{1}{r}(\cos\phi)\frac{\partial}{\partial\phi}\left[\frac{ik_t}{2k}\left(\{i\alpha-\beta\}e^{-i\phi}J_{\ell-1}+\{i\alpha+\beta\}e^{i\phi}J_{\ell+1}\right)\right]B_0\,e^{i(k_z z+\ell\phi-\omega t)}$$

$$=\hat{x}\frac{1}{r}(\cos\phi)\left[\frac{ik_t}{2k}\left(i(\ell-1)\{i\alpha-\beta\}e^{-i\phi}J_{\ell-1}+i(\ell+1)\{i\alpha+\beta\}e^{i\phi}J_{\ell+1}\right)\right]B_0\,e^{i(k_z z+\ell\phi-\omega t)}$$

$$=\hat{x}\frac{1}{r}(\cos\phi)\left[\frac{ik_t}{2k}\left(i(\ell-1)\{i\alpha-\beta\}e^{-i\phi}\left(\frac{k_t r\{J_\ell+J_{\ell-2}\}}{2(\ell-1)}\right)+i(\ell+1)\{i\alpha+\beta\}e^{i\phi}\left(\frac{k_t r\{J_\ell+J_{\ell+2}\}}{2(\ell+1)}\right)\right)\right]$$
$$\times B_0\,e^{i(k_z z+\ell\phi-\omega t)}$$

$$=\hat{x}\left[\frac{ik_t}{2k}\left(\{i\alpha-\beta\}e^{-i\phi}\left(\frac{k_t i(\cos\phi)\{J_\ell+J_{\ell-2}\}}{2}\right)+\{i\alpha+\beta\}e^{i\phi}\left(\frac{k_t i(\cos\phi)\{J_\ell+J_{\ell+2}\}}{2}\right)\right)\right]B_0\,e^{i(k_z z+\ell\phi-\omega t)}$$

$$=\hat{x}\left[\frac{ik_t^2}{4k}\left(\{i\alpha-\beta\}e^{-i\phi}i(\cos\phi)\{J_\ell+J_{\ell-2}\}+\{i\alpha+\beta\}e^{i\phi}i(\cos\phi)\{J_\ell+J_{\ell+2}\}\right)\right]B_0\,e^{i(k_z z+\ell\phi-\omega t)}.\qquad(1.20)$$

We made use of the Bessel relation $2\nu J_\nu(x)=x(J_{\nu+1}(x)+J_{\nu-1}(x))$, i.e. $2(\ell-1)J_{\ell-1}(k_t r)=k_t r\{J_\ell(k_t r)+J_{\ell-2}(k_t r)\}$. Adding the two *x* components (1.19) and (1.20) together yields

$$\hat{x}\frac{ik_t^2}{4k}\left[\begin{array}{l}((\sin\phi)\{i\alpha-\beta\}e^{-i\phi}\{(J_{\ell-2}-J_\ell)\}+(\sin\phi)\{i\alpha+\beta\}e^{i\phi}\{(J_\ell-J_{\ell+2})\})\\+(\{i\alpha-\beta\}e^{-i\phi}i(\cos\phi)\{J_\ell+J_{\ell-2}\}+\{i\alpha+\beta\}e^{i\phi}i(\cos\phi)\{J_\ell+J_{\ell+2}\})\end{array}\right]B_0\,e^{i(k_z z+\ell\phi-\omega t)}$$

$$=\hat{x}\frac{ik_t^2}{4k}\left[\begin{array}{l}-J_\ell(\sin\phi)\{i\alpha-\beta\}e^{-i\phi}+J_\ell i(\cos\phi)\{i\alpha-\beta\}e^{-i\phi}\\+J_\ell(\sin\phi)\{i\alpha+\beta\}e^{i\phi}+J_\ell i(\cos\phi)\{i\alpha+\beta\}e^{i\phi}\\+J_{\ell-2}(\sin\phi)\{i\alpha-\beta\}e^{-i\phi}+J_{\ell-2}i(\cos\phi)\{i\alpha-\beta\}e^{-i\phi}\\-J_{\ell+2}(\sin\phi)\{i\alpha+\beta\}e^{i\phi}+J_{\ell+2}i(\cos\phi)\{i\alpha+\beta\}e^{i\phi}\end{array}\right]B_0\,e^{i(k_z z+\ell\phi-\omega t)}$$

$$=\hat{x}\frac{ik_t^2}{4k}\left[\begin{array}{l}J_\ell i\{i\alpha-\beta\}e^{i\phi}e^{-i\phi}+J_\ell i\{i\alpha+\beta\}e^{i\phi}e^{-i\phi}\\+J_{\ell-2}i\{i\alpha-\beta\}e^{-i\phi}e^{-i\phi}+J_{\ell+2}i\{i\alpha+\beta\}e^{i\phi}e^{i\phi}\end{array}\right]B_0\,e^{i(k_z z+\ell\phi-\omega t)}$$

$$=\hat{x}\frac{ik_t^2}{4k}\left[\begin{array}{l}J_\ell\{-\alpha-i\beta\}+J_\ell\{-\alpha+i\beta\}\\+J_{\ell-2}\{-\alpha-i\beta\}e^{-2i\phi}+J_{\ell+2}\{i\beta-\alpha\}e^{2i\phi}\end{array}\right]B_0\,e^{i(k_z z+\ell\phi-\omega t)}$$

$$=\hat{x}\frac{ik_t^2}{4k}\left[-2\alpha J_\ell+J_{\ell-2}\{-\alpha-i\beta\}e^{-2i\phi}+J_{\ell+2}\{i\beta-\alpha\}e^{2i\phi}\right]B_0\,e^{i(k_z z+\ell\phi-\omega t)}.\qquad(1.21)$$

The *y* components (labelled 3 and 4 in (1.18)) are determined in a similar fashion:

$$-\hat{\mathbf{y}}(\cos\phi)\frac{\partial}{\partial r}\left[\frac{ik_t}{2k}\left(\{i\alpha-\beta\}e^{-i\phi}J_{\ell-1}+\{i\alpha+\beta\}e^{i\phi}J_{\ell+1}\right)\right]B_0\,e^{i(k_z z+\ell\phi-\omega t)}$$

$$=-\hat{\mathbf{y}}(\cos\phi)\left[\frac{ik_t}{2k}\left(\{i\alpha-\beta\}e^{-i\phi}\left\{\frac{k_t}{2}(J_{\ell-2}-J_\ell)\right\}+\{i\alpha+\beta\}e^{i\phi}\left\{\frac{k_t}{2}(J_\ell-J_{\ell+2})\right\}\right)\right]B_0\,e^{i(k_z z+\ell\phi-\omega t)}, \quad (1.22)$$

and

$$\hat{\mathbf{y}}\frac{1}{r}(\sin\phi)\frac{\partial}{\partial\phi}\left[\frac{ik_t}{2k}\left(\{i\alpha-\beta\}e^{-i\phi}J_{\ell-1}+\{i\alpha+\beta\}e^{i\phi}J_{\ell+1}\right)\right]B_0\,e^{i(k_z z+\ell\phi-\omega t)}$$

$$=\hat{\mathbf{y}}\frac{1}{r}(\sin\phi)\left[\frac{ik_t}{2k}\left(i(\ell-1)\{i\alpha-\beta\}e^{-i\phi}J_{\ell-1}+i(\ell+1)\{i\alpha+\beta\}e^{i\phi}J_{\ell+1}\right)\right]B_0\,e^{i(k_z z+\ell\phi-\omega t)}$$

$$=\hat{\mathbf{y}}\frac{1}{r}(\sin\phi)\left[\frac{ik_t}{2k}\left(i(\ell-1)\{i\alpha-\beta\}e^{-i\phi}\left(\frac{k_t r\{J_\ell+J_{\ell-2}\}}{2(\ell-1)}\right)+i(\ell+1)\{i\alpha+\beta\}e^{i\phi}\left(\frac{k_t r\{J_\ell+J_{\ell+2}\}}{2(\ell+1)}\right)\right)\right]$$
$$\times B_0\,e^{i(k_z z+\ell\phi-\omega t)}$$

$$=\hat{\mathbf{y}}\left[\frac{ik_t^2}{4k}\left(i\{i\alpha-\beta\}(\sin\phi)e^{-i\phi}\{J_\ell+J_{\ell-2}\}+i\{i\alpha+\beta\}(\sin\phi)e^{i\phi}\{J_\ell+J_{\ell+2}\}\right)\right]B_0\,e^{i(k_z z+\ell\phi-\omega t)}. \quad (1.23)$$

Adding (1.22) and (1.23) gives the total *y* component:

$$\hat{\mathbf{y}}\frac{ik_t^2}{4k}\begin{bmatrix}(\{i\alpha-\beta\}(-\cos\phi)e^{-i\phi}(J_{\ell-2}-J_\ell)+\{i\alpha+\beta\}(-\cos\phi)e^{i\phi}(J_\ell-J_{\ell+2}))\\(i\{i\alpha-\beta\}(\sin\phi)e^{-i\phi}\{J_\ell+J_{\ell-2}\}+i\{i\alpha+\beta\}(\sin\phi)e^{i\phi}\{J_\ell+J_{\ell+2}\})\end{bmatrix}B_0\,e^{i(k_z z+\ell\phi-\omega t)}$$

$$=\hat{\mathbf{y}}\frac{ik_t^2}{4k}\begin{bmatrix}J_\ell\{i\alpha-\beta\}(\cos\phi)e^{-i\phi}+J_\ell\{i\alpha-\beta\}i(\sin\phi)e^{-i\phi}\\+J_\ell\{i\alpha+\beta\}(-\cos\phi)e^{i\phi}+J_\ell\{i\alpha+\beta\}(i\sin\phi)e^{i\phi}\\+J_{\ell-2}\{i\alpha-\beta\}(-\cos\phi)e^{-i\phi}+J_{\ell-2}\{i\alpha-\beta\}(i\sin\phi)e^{-i\phi}\\-J_{\ell+2}\{i\alpha+\beta\}(-\cos\phi)e^{i\phi}+J_{\ell+2}\{i\alpha+\beta\}(i\sin\phi)e^{i\phi}\end{bmatrix}B_0\,e^{i(k_z z+\ell\phi-\omega t)}$$

$$=\hat{\mathbf{y}}\frac{ik_t^2}{4k}\begin{bmatrix}J_\ell\{i\alpha-\beta\}e^{i\phi}e^{-i\phi}-J_\ell\{i\alpha+\beta\}e^{-i\phi}e^{i\phi}\\-J_{\ell-2}\{i\alpha-\beta\}e^{-i\phi}e^{-i\phi}+J_{\ell+2}\{i\alpha+\beta\}e^{i\phi}e^{i\phi}\end{bmatrix}B_0\,e^{i(k_z z+\ell\phi-\omega t)}$$

$$=\hat{\mathbf{y}}\frac{ik_t^2}{4k}\left[J_\ell\{i\alpha-\beta\}-J_\ell\{i\alpha+\beta\}+J_{\ell-2}\{\beta-i\alpha\}e^{-2i\phi}+J_{\ell+2}\{i\alpha+\beta\}e^{2i\phi}\right]B_0\,e^{i(k_z z+\ell\phi-\omega t)}$$

$$=\hat{\mathbf{y}}\frac{ik_t^2}{4k}\left[-2\beta J_\ell+J_{\ell-2}\{\beta-i\alpha\}e^{-2i\phi}+J_{\ell+2}\{i\alpha+\beta\}e^{2i\phi}\right]B_0\,e^{i(k_z z+\ell\phi-\omega t)}. \quad (1.24)$$

Taking our total *x* (1.21) and *y* components (1.24) and placing them back into (1.18) gives the total electric second-order transverse field as:

$$\boldsymbol{E}^{(T_2)} = \begin{bmatrix} \hat{\boldsymbol{x}} \dfrac{k_t^2}{4k^2} \left[ 2\alpha J_\ell + J_{\ell-2}\{\alpha+i\beta\}e^{-2i\phi} + J_{\ell+2}\{\alpha-i\beta\}e^{2i\phi} \right] \\ +\hat{\boldsymbol{y}} \dfrac{k_t^2}{4k^2} \left[ 2\beta J_\ell + J_{\ell-2}\{i\alpha-\beta\}e^{-2i\phi} + J_{\ell+2}\{-\beta-i\alpha\}e^{2i\phi} \right] \end{bmatrix} E_0\, e^{i(k_z z + \ell\phi - \omega t)}. \qquad (1.25)$$

When added to the zeroth-order transverse and first-order longitudinal components (1.14) give the total electric field up to second order as

$$\boldsymbol{E} = \begin{bmatrix} (\alpha\hat{\boldsymbol{x}} + \beta\hat{\boldsymbol{y}}) J_\ell + \hat{\boldsymbol{z}} \dfrac{ik_t}{2k_z} \left( \{\alpha+i\beta\}e^{-i\phi} J_{\ell-1} + \{i\beta-\alpha\}e^{i\phi} J_{\ell+1} \right) \\ +\hat{\boldsymbol{x}} \dfrac{k_t^2}{4k^2} \left[ 2\alpha J_\ell + J_{\ell-2}\{\alpha+i\beta\}e^{-2i\phi} + J_{\ell+2}\{\alpha-i\beta\}e^{2i\phi} \right] \\ +\hat{\boldsymbol{y}} \dfrac{k_t^2}{4k^2} \left[ 2\beta J_\ell + J_{\ell-2}\{i\alpha-\beta\}e^{-2i\phi} + J_{\ell+2}\{-\beta-i\alpha\}e^{2i\phi} \right] \end{bmatrix} E_0\, e^{i(k_z z + \ell\phi - \omega t)}, \qquad (1.26)$$

as found in the main text. Carrying out an analogous procedure using Faraday's Law $\nabla \times \boldsymbol{E} = -\dfrac{\partial \boldsymbol{B}}{\partial t}$ provides us with the second-order transverse magnetic fields:

$$\begin{aligned} \boldsymbol{B} &= -\int (\nabla \times \boldsymbol{E})\partial t \\ &= -\dfrac{i}{\omega}\left( \hat{\boldsymbol{x}}\dfrac{\partial}{\partial x} + \hat{\boldsymbol{y}}\dfrac{\partial}{\partial y} \right)\left[ (\alpha\hat{\boldsymbol{x}} + \beta\hat{\boldsymbol{y}}) J_\ell + \hat{\boldsymbol{z}} \dfrac{ik_t}{2k_z}\left( \{i\beta+\alpha\}e^{-i\phi} J_{\ell-1} + \{i\beta-\alpha\}e^{i\phi} J_{\ell+1} \right) \right] E_0\, e^{i(k_z z + \ell\phi - \omega t)}. \end{aligned} \qquad (1.27)$$

Once again, like the electric field, we only need to concentrate on the transverse gradient of the longitudinal field to produce the second-order transverse fields. The calculations are given below without commentary.

$$-\frac{i}{\omega}\left(\hat{\boldsymbol{x}}\frac{\partial}{\partial x}+\hat{\boldsymbol{y}}\frac{\partial}{\partial y}\right)\left[\hat{\boldsymbol{z}}\frac{ik_t}{2k_z}\left(\{i\beta+\alpha\}\mathrm{e}^{-i\phi}J_{\ell-1}+\{i\beta-\alpha\}\mathrm{e}^{i\phi}J_{\ell+1}\right)\right]E_0\,\mathrm{e}^{i(k_z z+\ell\phi-\omega t)}$$

$$=\frac{i}{k}\left(\hat{\boldsymbol{y}}\frac{\partial}{\partial x}-\hat{\boldsymbol{x}}\frac{\partial}{\partial y}\right)\left[\frac{ik_t}{2k_z}\left(\{i\beta+\alpha\}\mathrm{e}^{-i\phi}J_{\ell-1}+\{i\beta-\alpha\}\mathrm{e}^{i\phi}J_{\ell+1}\right)\right]B_0\,\mathrm{e}^{i(k_z z+\ell\phi-\omega t)}$$

$$=\frac{i}{k}\left(\hat{\boldsymbol{y}}\left\{(\cos\phi)\frac{\partial}{\partial r}-\frac{1}{r}(\sin\phi)\frac{\partial}{\partial\phi}\right\}-\hat{\boldsymbol{x}}\left\{(\sin\phi)\frac{\partial}{\partial r}+\frac{1}{r}(\cos\phi)\frac{\partial}{\partial\phi}\right\}\right)$$
$$\times\left[\frac{ik_t}{2k_z}\left(\{i\beta+\alpha\}\mathrm{e}^{-i\phi}J_{\ell-1}+\{i\beta-\alpha\}\mathrm{e}^{i\phi}J_{\ell+1}\right)\right]B_0\,\mathrm{e}^{i(k_z z+\ell\phi-\omega t)}$$

$$=\frac{i}{k}\hat{\boldsymbol{y}}(\cos\phi)\frac{\partial}{\partial r}\underbrace{\left[\frac{ik_t}{2k_z}\left(\{i\beta+\alpha\}\mathrm{e}^{-i\phi}J_{\ell-1}+\{i\beta-\alpha\}\mathrm{e}^{i\phi}J_{\ell+1}\right)\right]B_0\,\mathrm{e}^{i(k_z z+\ell\phi-\omega t)}}_{1}$$

$$-\frac{i}{k}\hat{\boldsymbol{y}}\frac{1}{r}(\sin\phi)\frac{\partial}{\partial\phi}\underbrace{\left[\frac{ik_t}{2k_z}\left(\{i\beta+\alpha\}\mathrm{e}^{-i\phi}J_{\ell-1}+\{i\beta-\alpha\}\mathrm{e}^{i\phi}J_{\ell+1}\right)\right]B_0\,\mathrm{e}^{i(k_z z+\ell\phi-\omega t)}}_{2}$$

$$-\frac{i}{k}\hat{\boldsymbol{x}}(\sin\phi)\frac{\partial}{\partial r}\underbrace{\left[\frac{ik_t}{2k_z}\left(\{i\beta+\alpha\}\mathrm{e}^{-i\phi}J_{\ell-1}+\{i\beta-\alpha\}\mathrm{e}^{i\phi}J_{\ell+1}\right)\right]B_0\,\mathrm{e}^{i(k_z z+\ell\phi-\omega t)}}_{3}$$

$$-\frac{i}{k}\hat{\boldsymbol{x}}\frac{1}{r}(\cos\phi)\frac{\partial}{\partial\phi}\underbrace{\left[\frac{ik_t}{2k_z}\left(\{i\beta+\alpha\}\mathrm{e}^{-i\phi}J_{\ell-1}+\{i\beta-\alpha\}\mathrm{e}^{i\phi}J_{\ell+1}\right)\right]B_0\,\mathrm{e}^{i(k_z z+\ell\phi-\omega t)}}_{4}, \qquad (1.28)$$

$$\frac{i}{k}\left(\hat{\boldsymbol{y}}(\cos\phi)\frac{\partial}{\partial r}\right)\left[\frac{ik_t}{2k_z}\left(\{i\beta+\alpha\}\mathrm{e}^{-i\phi}J_{\ell-1}+\{i\beta-\alpha\}\mathrm{e}^{i\phi}J_{\ell+1}\right)\right]B_0\,\mathrm{e}^{i(k_z z+\ell\phi-\omega t)}$$

$$=\frac{i}{k}\left(\hat{\boldsymbol{y}}(\cos\phi)\right)\left[\frac{ik_t}{2k_z}\left(\{i\beta+\alpha\}\mathrm{e}^{-i\phi}\left\{\frac{k_t}{2}(J_{\ell-2}-J_\ell)\right\}+\{i\beta-\alpha\}\mathrm{e}^{i\phi}\left\{\frac{k_t}{2}(J_\ell-J_{\ell+2})\right\}\right)\right]B_0\,\mathrm{e}^{i(k_z z+\ell\phi-\omega t)}$$

$$=\frac{i}{k}\hat{\boldsymbol{y}}\left[\frac{ik_t^2}{4k_z}\left(\{i\beta+\alpha\}(\cos\phi)\mathrm{e}^{-i\phi}(J_{\ell-2}-J_\ell)+\{i\beta-\alpha\}(\cos\phi)\mathrm{e}^{i\phi}(J_\ell-J_{\ell+2})\right)\right]B_0\,\mathrm{e}^{i(k_z z+\ell\phi-\omega t)}, \qquad (1.29)$$

$$-\frac{i}{k}\hat{y}\frac{1}{r}(\sin\phi)\frac{\partial}{\partial\phi}\left[\frac{ik_t}{2k_z}\left(\{i\beta+\alpha\}e^{-i\phi}J_{\ell-1}+\{i\beta-\alpha\}e^{i\phi}J_{\ell+1}\right)\right]B_0\,e^{i(k_z z+\ell\phi-\omega t)}$$

$$=-\frac{i}{k}\hat{y}\frac{1}{r}\left[\frac{ik_t}{2k_z}\left(i(\ell-1)\{i\beta+\alpha\}(\sin\phi)e^{-i\phi}J_{\ell-1}+i(\ell+1)\{i\beta-\alpha\}(\sin\phi)e^{i\phi}J_{\ell+1}\right)\right]B_0\,e^{i(k_z z+\ell\phi-\omega t)}$$

$$=-\frac{i}{k}\hat{y}\frac{1}{r}\left[\frac{ik_t}{2k_z}\begin{pmatrix}i(\ell-1)\{i\beta+\alpha\}(\sin\phi)e^{-i\phi}\left(\dfrac{k_t r\{J_\ell+J_{\ell-2}\}}{2(\ell-1)}\right)\\+i(\ell+1)\{i\beta-\alpha\}(\sin\phi)e^{i\phi}\left(\dfrac{k_t r\{J_\ell+J_{\ell+2}\}}{2(\ell+1)}\right)\end{pmatrix}\right]B_0\,e^{i(k_z z+\ell\phi-\omega t)}$$

$$=-\frac{i}{k}\hat{y}\left[\frac{ik_t^2}{4k_z}\left(\{i\beta+\alpha\}(i\sin\phi)e^{-i\phi}\{J_\ell+J_{\ell-2}\}+\{i\beta-\alpha\}(i\sin\phi)e^{i\phi}\{J_\ell+J_{\ell+2}\}\right)\right]B_0\,e^{i(k_z z+\ell\phi-\omega t)}, \quad (1.30)$$

The sum of (1.29) and (1.30)

$$\frac{i}{k}\hat{y}\left[\frac{ik_t^2}{4k_z}\left(\{i\beta+\alpha\}(\cos\phi)e^{-i\phi}(J_{\ell-2}-J_\ell)+\{i\beta-\alpha\}(\cos\phi)e^{i\phi}(J_\ell-J_{\ell+2})\right)\right]B_0\,e^{i(k_z z+\ell\phi-\omega t)}$$

$$-\frac{i}{k}\hat{y}\left[\frac{ik_t^2}{4k_z}\left(\{i\beta+\alpha\}(i\sin\phi)e^{-i\phi}\{J_\ell+J_{\ell-2}\}+\{i\beta-\alpha\}(i\sin\phi)e^{i\phi}\{J_\ell+J_{\ell+2}\}\right)\right]B_0\,e^{i(k_z z+\ell\phi-\omega t)}$$

$$=\hat{y}\frac{k_t^2}{4kk_z}\left[\begin{array}{l}\left(\{i\beta+\alpha\}(i\sin\phi)e^{-i\phi}\{J_\ell+J_{\ell-2}\}+\{i\beta-\alpha\}(i\sin\phi)e^{i\phi}\{J_\ell+J_{\ell+2}\}\right)\\-\left(\{i\beta+\alpha\}(\cos\phi)e^{-i\phi}(J_{\ell-2}-J_\ell)+\{i\beta-\alpha\}(\cos\phi)e^{i\phi}(J_\ell-J_{\ell+2})\right)\end{array}\right]B_0\,e^{i(k_z z+\ell\phi-\omega t)}$$

$$=\hat{y}\frac{k_t^2}{4kk_z}\left[\begin{array}{l}J_\ell\{i\beta+\alpha\}(i\sin\phi)e^{-i\phi}+J_\ell\{i\beta+\alpha\}(\cos\phi)e^{-i\phi}\\+J_\ell\{i\beta-\alpha\}(i\sin\phi)e^{i\phi}-J_\ell\{i\beta-\alpha\}(\cos\phi)e^{i\phi}\\+J_{\ell-2}\{i\beta+\alpha\}(i\sin\phi)e^{-i\phi}-J_{\ell-2}\{i\beta+\alpha\}(\cos\phi)e^{-i\phi}\\+J_{\ell+2}\{i\beta-\alpha\}(i\sin\phi)e^{i\phi}+J_{\ell+2}\{i\beta-\alpha\}(\cos\phi)e^{i\phi}\end{array}\right]B_0\,e^{i(k_z z+\ell\phi-\omega t)}$$

$$=\hat{y}\frac{k_t^2}{4kk_z}\left[\begin{array}{l}J_\ell\{i\beta+\alpha\}e^{i\phi}e^{-i\phi}-J_\ell\{i\beta-\alpha\}e^{-i\phi}e^{i\phi}\\-J_{\ell-2}\{i\beta+\alpha\}e^{-i\phi}e^{-i\phi}+J_{\ell+2}\{i\beta-\alpha\}e^{i\phi}e^{i\phi}\end{array}\right]B_0\,e^{i(k_z z+\ell\phi-\omega t)}$$

$$=\hat{y}\frac{k_t^2}{4kk_z}\left[J_\ell\{i\beta+\alpha\}-J_\ell\{i\beta-\alpha\}+J_{\ell-2}\{-i\beta-\alpha\}e^{-2i\phi}+J_{\ell+2}\{i\beta-\alpha\}e^{2i\phi}\right]B_0\,e^{i(k_z z+\ell\phi-\omega t)}$$

$$=\hat{y}\frac{k_t^2}{4kk_z}\left[2\alpha J_\ell+J_{\ell-2}\{-i\beta-\alpha\}e^{-2i\phi}+J_{\ell+2}\{i\beta-\alpha\}e^{2i\phi}\right]B_0\,e^{i(k_z z+\ell\phi-\omega t)}. \quad (1.31)$$

$$-\frac{i}{k}\left(\hat{x}(\sin\phi)\frac{\partial}{\partial r}\right)\left[\frac{ik_t}{2k_z}\left(\{i\beta+\alpha\}e^{-i\phi}J_{\ell-1}+\{i\beta-\alpha\}e^{i\phi}J_{\ell+1}\right)\right]B_0\,e^{i(k_z z+\ell\phi-\omega t)}$$

$$=-\frac{i}{k}\hat{x}\left[\frac{ik_t}{2k_z}\left(\{i\beta+\alpha\}(\sin\phi)e^{-i\phi}\left\{\frac{k_t}{2}(J_{\ell-2}-J_\ell)\right\}+\{i\beta-\alpha\}(\sin\phi)e^{i\phi}\left\{\frac{k_t}{2}(J_\ell-J_{\ell+2})\right\}\right)\right]B_0\,e^{i(k_z z+\ell\phi-\omega t)}$$

$$=\frac{k_t^2}{4kk_z}\hat{x}\left[\left(\{i\beta+\alpha\}(\sin\phi)e^{-i\phi}(J_{\ell-2}-J_\ell)+\{i\beta-\alpha\}(\sin\phi)e^{i\phi}(J_\ell-J_{\ell+2})\right)\right]B_0\,e^{i(k_z z+\ell\phi-\omega t)}, \qquad (1.32)$$

$$-\frac{i}{k}\left(\hat{x}\frac{1}{r}(\cos\phi)\frac{\partial}{\partial\phi}\right)\left[\frac{ik_t}{2k_z}\left(\{i\beta+\alpha\}e^{-i\phi}J_{\ell-1}+\{i\beta-\alpha\}e^{i\phi}J_{\ell+1}\right)\right]B_0\,e^{i(k_z z+\ell\phi-\omega t)}$$

$$=-\frac{i}{k}\left(\hat{x}\frac{1}{r}(\cos\phi)\right)\frac{ik_t}{2k_z}\left[\begin{array}{l}i(\ell-1)\{i\beta+\alpha\}e^{-i\phi}\left(\dfrac{k_t r\{J_\ell+J_{\ell-2}\}}{2(\ell-1)}\right)\\+i(\ell+1)\{i\beta-\alpha\}e^{i\phi}\left(\dfrac{k_t r\{J_\ell+J_{\ell+2}\}}{2(\ell+1)}\right)\end{array}\right]B_0\,e^{i(k_z z+\ell\phi-\omega t)}$$

$$=-\frac{i}{k}\left(\hat{x}(\cos\phi)\right)\left[\frac{ik_t^2}{4k_z}\left(i\{i\beta+\alpha\}e^{-i\phi}\{J_\ell+J_{\ell-2}\}+i\{i\beta-\alpha\}e^{i\phi}\{J_\ell+J_{\ell+2}\}\right)\right]B_0\,e^{i(k_z z+\ell\phi-\omega t)}$$

$$=\frac{k_t^2}{4kk_z}\hat{x}\left[\left(i\{i\beta+\alpha\}(\cos\phi)e^{-i\phi}\{J_\ell+J_{\ell-2}\}+i\{i\beta-\alpha\}(\cos\phi)e^{i\phi}\{J_\ell+J_{\ell+2}\}\right)\right]B_0\,e^{i(k_z z+\ell\phi-\omega t)}, \qquad (1.33)$$

Adding (1.32) and (1.33) gives

$$\frac{k_t^2}{4kk_z}\hat{\boldsymbol{x}}\Big[\big(\{i\beta+\alpha\}(\sin\phi)\mathrm{e}^{-i\phi}(J_{\ell-2}-J_\ell)+\{i\beta-\alpha\}(\sin\phi)\mathrm{e}^{i\phi}(J_\ell-J_{\ell+2})\big)\Big]B_0\,\mathrm{e}^{i(k_z z+\ell\phi-\omega t)}$$

$$+\frac{k_t^2}{4kk_z}\hat{\boldsymbol{x}}\Big[\big(i\{i\beta+\alpha\}(\cos\phi)\mathrm{e}^{-i\phi}\{J_\ell+J_{\ell-2}\}+i\{i\beta-\alpha\}(\cos\phi)\mathrm{e}^{i\phi}\{J_\ell+J_{\ell+2}\}\big)\Big]B_0\,\mathrm{e}^{i(k_z z+\ell\phi-\omega t)}$$

$$=\frac{k_t^2}{4kk_z}\hat{\boldsymbol{x}}\begin{bmatrix}\big(\{i\beta+\alpha\}(\sin\phi)\mathrm{e}^{-i\phi}(J_{\ell-2}-J_\ell)+\{i\beta-\alpha\}(\sin\phi)\mathrm{e}^{i\phi}(J_\ell-J_{\ell+2})\big)\\+\big(i\{i\beta+\alpha\}(\cos\phi)\mathrm{e}^{-i\phi}\{J_\ell+J_{\ell-2}\}+i\{i\beta-\alpha\}(\cos\phi)\mathrm{e}^{i\phi}\{J_\ell+J_{\ell+2}\}\big)\end{bmatrix}B_0\,\mathrm{e}^{i(k_z z+\ell\phi-\omega t)}$$

$$=\frac{k_t^2}{4kk_z}\hat{\boldsymbol{x}}\begin{bmatrix}-J_\ell\{i\beta+\alpha\}(\sin\phi)\mathrm{e}^{-i\phi}+J_\ell\{i\beta+\alpha\}(i\cos\phi)\mathrm{e}^{-i\phi}\\+J_\ell\{i\beta-\alpha\}(\sin\phi)\mathrm{e}^{i\phi}+iJ_\ell\{i\beta-\alpha\}(\cos\phi)\mathrm{e}^{i\phi}\\+J_{\ell-2}\{i\beta+\alpha\}(\sin\phi)\mathrm{e}^{-i\phi}+iJ_{\ell-2}\{i\beta+\alpha\}(\cos\phi)\mathrm{e}^{-i\phi}\\-J_{\ell+2}\{i\beta-\alpha\}(\sin\phi)\mathrm{e}^{i\phi}+iJ_{\ell+2}\{i\beta-\alpha\}(\cos\phi)\mathrm{e}^{i\phi}\end{bmatrix}B_0\,\mathrm{e}^{i(k_z z+\ell\phi-\omega t)}$$

$$=\frac{k_t^2}{4kk_z}\hat{\boldsymbol{x}}\begin{bmatrix}J_\ell\{i\beta+\alpha\}i\mathrm{e}^{i\phi}\mathrm{e}^{-i\phi}+J_\ell\{i\beta-\alpha\}i\mathrm{e}^{-i\phi}\mathrm{e}^{i\phi}\\+J_{\ell-2}\{i\beta+\alpha\}i\mathrm{e}^{-i\phi}\mathrm{e}^{-i\phi}+J_{\ell+2}\{i\beta-\alpha\}i\mathrm{e}^{i\phi}\mathrm{e}^{i\phi}\end{bmatrix}B_0\,\mathrm{e}^{i(k_z z+\ell\phi-\omega t)}$$

$$=\frac{k_t^2}{4kk_z}\hat{\boldsymbol{x}}\begin{bmatrix}J_\ell\{i\beta+\alpha\}i+J_\ell\{i\beta-\alpha\}i\\+J_{\ell-2}\{i\beta+\alpha\}i\mathrm{e}^{-2i\phi}+J_{\ell+2}\{i\beta-\alpha\}i\mathrm{e}^{2i\phi}\end{bmatrix}B_0\,\mathrm{e}^{i(k_z z+\ell\phi-\omega t)}$$

$$=\frac{k_t^2}{4kk_z}\hat{\boldsymbol{x}}\begin{bmatrix}J_\ell\{-\beta+i\alpha\}+J_\ell\{-\beta-i\alpha\}\\+J_{\ell-2}\{i\alpha-\beta\}\mathrm{e}^{-2i\phi}+J_{\ell+2}\{-i\alpha-\beta\}\mathrm{e}^{2i\phi}\end{bmatrix}B_0\,\mathrm{e}^{i(k_z z+\ell\phi-\omega t)} \quad (1.34)$$

$$=\frac{k_t^2}{4kk_z}\hat{\boldsymbol{x}}\Big[-2\beta J_\ell+J_{\ell-2}\{i\alpha-\beta\}\mathrm{e}^{-2i\phi}+J_{\ell+2}\{-i\alpha-\beta\}\mathrm{e}^{2i\phi}\Big]B_0\,\mathrm{e}^{i(k_z z+\ell\phi-\omega t)}.$$

Therefore, the total magnetic field is

$$\boldsymbol{B}=\begin{bmatrix}(\alpha\hat{\boldsymbol{y}}-\beta\hat{\boldsymbol{x}})\dfrac{k_z}{k}J_\ell+\hat{\boldsymbol{z}}\dfrac{ik_t}{2k}\big(\{i\alpha-\beta\}\mathrm{e}^{-i\phi}J_{\ell-1}+\{i\alpha+\beta\}\mathrm{e}^{i\phi}J_{\ell+1}\big)\\+\dfrac{k_t^2}{4kk_z}\Big(\hat{\boldsymbol{x}}\big[-2\beta J_\ell+J_{\ell-2}\{i\alpha-\beta\}\mathrm{e}^{-2i\phi}+J_{\ell+2}\{-i\alpha-\beta\}\mathrm{e}^{2i\phi}\big]\\+\hat{\boldsymbol{y}}\big[2\alpha J_\ell+J_{\ell-2}\{-i\beta-\alpha\}\mathrm{e}^{-2i\phi}+J_{\ell+2}\{i\beta-\alpha\}\mathrm{e}^{2i\phi}\big]\Big)\end{bmatrix}B_0\,\mathrm{e}^{i(k_z z+\ell\phi-\omega t)}. \quad (1.35)$$

## Calculation of the Optical Chirality Density

The optical chirality density for a quasi-monochromatic beam is given by the following formula:

$$\begin{aligned}
C &= \frac{\varepsilon_0}{2}\left[ \boldsymbol{E} \cdot \nabla \times \boldsymbol{E} + c^2 \boldsymbol{B} \cdot \nabla \times \boldsymbol{B} \right] \\
&= \frac{\varepsilon_0}{2}\left[ \boldsymbol{E} \cdot \left(-\frac{\partial}{\partial t}\boldsymbol{B}\right) + c^2 \boldsymbol{B} \cdot \left(\frac{1}{c^2}\frac{\partial}{\partial t}\boldsymbol{E}\right) \right] \\
&= \frac{\varepsilon_0}{2}\left[ \begin{array}{l} \frac{1}{2}\left(\boldsymbol{E}\,\mathrm{e}^{-i\omega t} + \bar{\boldsymbol{E}}\,\mathrm{e}^{i\omega t}\right) \cdot \left(-\frac{1}{2}\frac{\partial}{\partial t}\left(\boldsymbol{B}\,\mathrm{e}^{-i\omega t} + \bar{\boldsymbol{B}}\,\mathrm{e}^{i\omega t}\right)\right) \\ + c^2 \frac{1}{2}\left(\boldsymbol{B}\,\mathrm{e}^{-i\omega t} + \bar{\boldsymbol{B}}\,\mathrm{e}^{i\omega t}\right) \cdot \left(\frac{1}{c^2}\frac{\partial}{\partial t}\frac{1}{2}\left(\boldsymbol{E}\,\mathrm{e}^{-i\omega t} + \bar{\boldsymbol{E}}\,\mathrm{e}^{i\omega t}\right)\right) \end{array} \right] \\
&= \frac{\varepsilon_0}{8}i\omega\left[ \left(\boldsymbol{E}\,\mathrm{e}^{-i\omega t} + \bar{\boldsymbol{E}}\,\mathrm{e}^{i\omega t}\right) \cdot \left(\boldsymbol{B}\,\mathrm{e}^{-i\omega t} - \bar{\boldsymbol{B}}\,\mathrm{e}^{i\omega t}\right) + \left(\boldsymbol{B}\,\mathrm{e}^{-i\omega t} + \bar{\boldsymbol{B}}\,\mathrm{e}^{i\omega t}\right) \cdot \left(\bar{\boldsymbol{E}}\,\mathrm{e}^{i\omega t} - \boldsymbol{E}\,\mathrm{e}^{-i\omega t}\right) \right] \quad (1.36)\\
&= \frac{\varepsilon_0}{8}i\omega\left[ \left(\boldsymbol{E}\cdot\boldsymbol{B}\,\mathrm{e}^{-2i\omega t} + \bar{\boldsymbol{E}}\cdot\boldsymbol{B} - \boldsymbol{E}\cdot\bar{\boldsymbol{B}} - \bar{\boldsymbol{E}}\cdot\bar{\boldsymbol{B}}\,\mathrm{e}^{2i\omega t}\right) + \left(\boldsymbol{B}\cdot\bar{\boldsymbol{E}} - \boldsymbol{B}\cdot\boldsymbol{E}\,\mathrm{e}^{-2i\omega t} + \bar{\boldsymbol{B}}\cdot\bar{\boldsymbol{E}}\,\mathrm{e}^{2i\omega t} - \bar{\boldsymbol{B}}\cdot\boldsymbol{E}\right) \right] \\
&= \frac{\varepsilon_0}{4}i\omega\left[ \bar{\boldsymbol{E}}\cdot\boldsymbol{B} - \boldsymbol{E}\cdot\bar{\boldsymbol{B}} \right] \\
&= -\frac{\varepsilon_0\omega}{2}\mathrm{Im}\left(\bar{\boldsymbol{E}}\cdot\boldsymbol{B}\right)
\end{aligned}$$

It is interesting to note that, unlike the energy density, we do not need to average (1.36) over a characteristic time scale, it is naturally time-independent. In the third line of (1.36) we use

$$\boldsymbol{E}[r,t] = \mathrm{Re}\left\{\boldsymbol{E}[r]\mathrm{e}^{-i\omega t}\right\} = \frac{1}{2}\left(\boldsymbol{E}[r]\mathrm{e}^{-i\omega t} + \bar{\boldsymbol{E}}[r]\mathrm{e}^{i\omega t}\right) \text{ and } \boldsymbol{B}[r,t] = \mathrm{Re}\left\{\boldsymbol{B}[r]\mathrm{e}^{-i\omega t}\right\} = \frac{1}{2}\left(\boldsymbol{B}[r]\mathrm{e}^{-i\omega t} + \bar{\boldsymbol{B}}[r]\mathrm{e}^{i\omega t}\right).$$

It is worthwhile to show the calculation for a circularly polarized plane wave:

$$\boldsymbol{E} = \frac{1}{\sqrt{2}}(\hat{\boldsymbol{x}} + i\sigma\hat{\boldsymbol{y}})E_0\,\mathrm{e}^{ikz-i\omega t} \text{ and } \boldsymbol{B} = \frac{1}{\sqrt{2}}(\hat{\boldsymbol{y}} - i\sigma\hat{\boldsymbol{x}})B_0\,\mathrm{e}^{ikz-i\omega t}. \quad (1.37)$$

Such that:

$$\begin{aligned}
C &= -\frac{\varepsilon_0\omega}{2}\mathrm{Im}\left(\bar{\boldsymbol{E}}\cdot\boldsymbol{B}\right) \\
&= -\frac{\varepsilon_0\omega}{2}\mathrm{Im}\left(\left(\frac{1}{\sqrt{2}}(\hat{\boldsymbol{x}} - i\sigma\hat{\boldsymbol{y}})E_0\,\mathrm{e}^{-ikz+i\omega t}\right)\cdot\left(\frac{1}{\sqrt{2}}(\hat{\boldsymbol{y}} - i\sigma\hat{\boldsymbol{x}})B_0\,\mathrm{e}^{ikz-i\omega t}\right)\right) \\
&= \frac{\varepsilon_0\omega}{2}E_0 B_0\sigma \\
&= \frac{\varepsilon_0\omega}{2c}E_0^2\sigma \\
&= \frac{I\omega\sigma}{c^2}. \quad (1.38)
\end{aligned}$$

Inserting the electric (1.26) and magnetic field (1.35) into (1.36) yields of the optical chirality density of the Bessel beam:

$$\begin{aligned}
C &= -\frac{\varepsilon_0 \omega}{2} \operatorname{Im}(\bar{\boldsymbol{E}} \cdot \boldsymbol{B}) \\
&= -\frac{\varepsilon_0 \omega}{2} \operatorname{Im}\Bigg[\Bigg\{(\bar{\alpha}\hat{\boldsymbol{x}} + \bar{\beta}\hat{\boldsymbol{y}})J_\ell - \hat{\boldsymbol{z}}\frac{ik_t}{2k_z}\left(\{\bar{\alpha}-i\bar{\beta}\}e^{i\phi}J_{\ell-1} + \{-i\bar{\beta}-\bar{\alpha}\}e^{-i\phi}J_{\ell+1}\right) \\
&\quad + \hat{\boldsymbol{x}}\frac{k_t^2}{4k^2}\left[2\bar{\alpha}J_\ell + J_{\ell-2}\{\bar{\alpha}-i\bar{\beta}\}e^{2i\phi} + J_{\ell+2}\{\bar{\alpha}+i\bar{\beta}\}e^{-2i\phi}\right] \\
&\quad + \hat{\boldsymbol{y}}\frac{k_t^2}{4k^2}\left[2\bar{\beta}J_\ell + J_{\ell-2}\{-i\bar{\alpha}-\bar{\beta}\}e^{2i\phi} + J_{\ell+2}\{-\bar{\beta}+i\bar{\alpha}\}e^{-2i\phi}\right]\Bigg\}E_0\, e^{-i(k_z z+\ell\phi)} \\
&\quad \cdot \Bigg\{(\alpha\hat{\boldsymbol{y}}-\beta\hat{\boldsymbol{x}})\frac{k_z}{k}J_\ell + \hat{\boldsymbol{z}}\frac{ik_t}{2k}\left(\{i\alpha-\beta\}e^{-i\phi}J_{\ell-1} + \{i\alpha+\beta\}e^{i\phi}J_{\ell+1}\right) \\
&\quad + \frac{k_t^2}{4kk_z}\Big(\hat{\boldsymbol{x}}\left[-2\beta J_\ell + J_{\ell-2}\{i\alpha-\beta\}e^{-2i\phi} + J_{\ell+2}\{-i\alpha-\beta\}e^{2i\phi}\right] \\
&\quad + \hat{\boldsymbol{y}}\left[2\alpha J_\ell + J_{\ell-2}\{-i\beta-\alpha\}e^{-2i\phi} + J_{\ell+2}\{i\beta-\alpha\}e^{2i\phi}\right]\Big)\Bigg\}B_0\, e^{i(k_z z+\ell\phi)}\Bigg].
\end{aligned}$$

(1.39)

We can split this calculation up into a distinct number of contributions: those which involve the zeroth-order transverse fields in the dot product and are zeroth order with respect to the smallness parameter; those which involve the first-order longitudinal fields and are second-order in the smallness parameter; those which involve the zeroth-order transverse fields and the second-order transverse fields and are second order in the smallness parameter; and finally those which involve second-order transverse electric field coupling to second-order magnetic field and are fourth order in the smallness parameter.

$$\begin{aligned}
C(\mathrm{T}_0 \cdot \mathrm{T}_0) &= -\frac{\varepsilon_0 \omega}{2}\operatorname{Im}(\bar{\alpha}\hat{\boldsymbol{x}}+\bar{\beta}\hat{\boldsymbol{y}})J_\ell E_0\, e^{-i(k_z z+\ell\phi)} \cdot (\alpha\hat{\boldsymbol{y}}-\beta\hat{\boldsymbol{x}})\frac{k_z}{k}J_\ell B_0\, e^{i(k_z z+\ell\phi)} \\
&= -\frac{\varepsilon_0 \omega}{2c}\frac{k_z}{k}E_0^2 J_\ell^2 \operatorname{Im}(\alpha\bar{\beta}-\bar{\alpha}\beta) \\
&= -\frac{I\omega}{c^2}\frac{k_z}{k}J_\ell^2 \operatorname{Im}(\alpha\bar{\beta}-\bar{\alpha}\beta).
\end{aligned}$$

(1.40)

To further generalise our results we use the following formulae $\alpha = \cos\eta\cos\theta - i\sin\eta\sin\theta$ and $\beta = -\cos\eta\sin\theta - i\sin\eta\cos\theta$ where $\theta$ is the azimuth (it tells us the orientation of linearly polarised light is, e.g. *x*-pol, *y*-pol, diagonally pol, etc.) and $\eta$ is the ellipticity. Using this we can show that

$$\alpha\bar{\beta} - \bar{\alpha}\beta = (\cos\eta\cos\theta - i\sin\eta\sin\theta)(-\cos\eta\sin\theta + i\sin\eta\cos\theta)$$
$$- (\cos\eta\cos\theta + i\sin\eta\sin\theta)(-\cos\eta\sin\theta - i\sin\eta\cos\theta)$$
$$= -\cos^2\eta\cos\theta\sin\theta + i\sin\eta\cos\eta\cos^2\theta + i\sin\eta\cos\eta\sin^2\theta$$
$$+ \sin^2\eta\sin\theta\cos\theta + \cos^2\eta\cos\theta\sin\theta + i\sin\eta\cos\eta\cos^2\theta$$
$$+ i\sin\eta\cos\eta\sin^2\theta - \sin^2\eta\cos\theta\sin\theta$$
$$= i\sin\eta\cos\eta\left(\cos^2\theta + \sin^2\theta\right) + i\sin\eta\cos\eta\left(\cos^2\theta + \sin^2\theta\right)$$
$$- \left(\cos^2\eta + \sin^2\eta\right)\cos\theta\sin\theta + \left(\sin^2\eta + \cos^2\eta\right)\sin\theta\cos\theta$$
$$= 2i\sin\eta\cos\eta$$
$$= i\sin 2\eta. \tag{1.41}$$

Such that (1.40) becomes:

$$C(T_0 \cdot T_0) = -\frac{I\omega}{c^2}\frac{k_z}{k}J_\ell^2 \sin 2\eta. \tag{1.42}$$

Now to calculate the longitudinal field contribution:

$$C(L_1 \cdot L_1) = -\frac{\varepsilon_0 \omega}{2} \operatorname{Im}\left[ -\hat{z}\frac{ik_t}{2k_z}\left(\{\bar{\alpha} - i\bar{\beta}\}e^{i\phi} J_{\ell-1} + \{-i\bar{\beta} - \bar{\alpha}\}e^{-i\phi} J_{\ell+1}\right)\right] E_0\, e^{-i(k_z z + \ell\phi)}$$
$$\cdot \left[\hat{z}\frac{ik_t}{2k}\left(\{i\alpha - \beta\}e^{-i\phi} J_{\ell-1} + \{i\alpha + \beta\}e^{i\phi} J_{\ell+1}\right)\right] B_0\, e^{i(k_z z + \ell\phi)}$$

$$= -\frac{\varepsilon_0 \omega}{2c} E_0^2 \frac{k_t^2}{4kk_z} \operatorname{Im}\left(\{\bar{\alpha} - i\bar{\beta}\}e^{i\phi} J_{\ell-1} + \{-i\bar{\beta} - \bar{\alpha}\}e^{-i\phi} J_{\ell+1}\right)\left(\{i\alpha - \beta\}e^{-i\phi} J_{\ell-1} + \{i\alpha + \beta\}e^{i\phi} J_{\ell+1}\right)$$

$$= -\frac{\varepsilon_0 \omega}{2c} E_0^2 \frac{k_t^2}{4kk_z} \operatorname{Im}\begin{pmatrix}\{i|\alpha|^2 + \alpha\bar{\beta} - \bar{\alpha}\beta + i|\beta|^2\}J_{\ell-1}^2 + \{\alpha\bar{\beta} - i|\alpha|^2 + i|\beta|^2 + \bar{\alpha}\beta\}e^{-2i\phi} J_{\ell-1}J_{\ell+1} \\ + \{i|\alpha|^2 + \alpha\bar{\beta} + \bar{\alpha}\beta - i|\beta|^2\}e^{2i\phi} J_{\ell-1}J_{\ell+1} + \{\alpha\bar{\beta} - i|\alpha|^2 - i|\beta|^2 - \bar{\alpha}\beta\}J_{\ell+1}^2\end{pmatrix}$$

$$= -\frac{\varepsilon_0 \omega}{2c} E_0^2 \frac{k_t^2}{4kk_z} \operatorname{Im}\begin{pmatrix}\{\alpha\bar{\beta} - \bar{\alpha}\beta + i\}J_{\ell-1}^2 + \{\alpha\bar{\beta} - i|\alpha|^2 + i|\beta|^2 + \bar{\alpha}\beta\}e^{-2i\phi} J_{\ell-1}J_{\ell+1} \\ + \{i|\alpha|^2 + \alpha\bar{\beta} + \bar{\alpha}\beta - i|\beta|^2\}e^{2i\phi} J_{\ell-1}J_{\ell+1} + \{\alpha\bar{\beta} - \bar{\alpha}\beta - 1\}J_{\ell+1}^2\end{pmatrix}$$

$$= -\frac{\varepsilon_0 \omega}{2c} E_0^2 \frac{k_t^2}{4kk_z} \operatorname{Im}\left(\{\alpha\bar{\beta} - \bar{\alpha}\beta + i\}J_{\ell-1}^2 + \{\alpha\bar{\beta} - \bar{\alpha}\beta - i\}J_{\ell+1}^2\right)$$

$$= -\frac{I\omega}{c^2}\frac{k_t^2}{4kk_z}\left(\{\sin 2\eta + 1\}J_{\ell-1}^2 + \{\sin 2\eta - 1\}J_{\ell+1}^2\right). \tag{1.43}$$

The final step is due to the fact that $\text{Im}\{\alpha\bar{\beta} - i|\alpha|^2 + i|\beta|^2 + \bar{\alpha}\beta\}e^{-2i\phi}J_{\ell-1}J_{\ell+1}$ and $\text{Im}\{i|\alpha|^2 + \alpha\bar{\beta} + \bar{\alpha}\beta - i|\beta|^2\}e^{2i\phi}J_{\ell-1}J_{\ell+1}$ are zero for any value of $\alpha$ and $\beta$.

Now we calculate the mixed terms in the cross product between the zeroth order and second order transverse fields. It is important to include all relevant terms: zeroth-order electric dot product with second order magnetic; zeroth order magnetic with second-order electric:

$$C(T_0 \cdot T_2) = -\frac{\varepsilon_0 \omega}{2}\text{Im}\left[(\bar{\alpha}\hat{x} + \bar{\beta}\hat{y})J_\ell E_0\, e^{-i(k_z z + \ell\phi - \omega t)} \cdot \frac{k_t^2}{4kk_z}\left(\hat{x}\left[-2\beta J_\ell + J_{\ell-2}\{i\alpha - \beta\}e^{-2i\phi} + J_{\ell+2}\{-i\alpha - \beta\}e^{2i\phi}\right]\right.\right.$$

$$+\hat{y}\left[2\alpha J_\ell + J_{\ell-2}\{-i\beta - \alpha\}e^{-2i\phi} + J_{\ell+2}\{i\beta - \alpha\}e^{2i\phi}\right]\right)B_0\, e^{i(k_z z + \ell\phi)}$$

$$+(\alpha\hat{y} - \beta\hat{x})\frac{k_z}{k}J_\ell B_0\, e^{i(k_z z + \ell\phi - \omega t)} \cdot \frac{k_t^2}{4k^2}\left(\hat{x}\left[2\bar{\alpha}J_\ell + J_{\ell-2}\{\bar{\alpha} - i\bar{\beta}\}e^{2i\phi} + J_{\ell+2}\{\bar{\alpha} + i\bar{\beta}\}e^{-2i\phi}\right]\right.$$

$$\left.\left.+\hat{y}\left[2\bar{\beta}J_\ell + J_{\ell-2}\{-i\bar{\alpha} - \bar{\beta}\}e^{2i\phi} + J_{\ell+2}\{-\bar{\beta} + i\bar{\alpha}\}e^{-2i\phi}\right]\right)E_0\, e^{-i(k_z z + \ell\phi)}\right]$$

$$= -\frac{\varepsilon_0 \omega}{2c}E_0^2\, \text{Im}\left[\frac{k_t^2}{4kk_z}\left\{-2\bar{\alpha}\beta J_\ell^2 + J_\ell J_{\ell-2}\{i|\alpha|^2 - \bar{\alpha}\beta\}e^{-2i\phi} + J_\ell J_{\ell+2}\{-i|\alpha|^2 - \bar{\alpha}\beta\}e^{2i\phi}\right.\right.$$

$$\left.+2\alpha\bar{\beta}J_\ell^2 + J_\ell J_{\ell-2}\{-i|\beta|^2 - \alpha\bar{\beta}\}e^{-2i\phi} + J_\ell J_{\ell+2}\{i|\beta|^2 - \alpha\bar{\beta}\}e^{2i\phi}\right)$$

$$+\frac{k_z k_t^2}{4k^3}\left\{2\alpha\bar{\beta}J_\ell^2 + J_\ell J_{\ell-2}\{-i|\alpha|^2 - \alpha\bar{\beta}\}e^{2i\phi} + J_\ell J_{\ell+2}\{-\alpha\bar{\beta} + i|\alpha|^2\}e^{-2i\phi}\right.$$

$$\left.\left.-2\bar{\alpha}\beta J_\ell^2 + J_\ell J_{\ell-2}\{-\bar{\alpha}\beta + i|\beta|^2\}e^{2i\phi} + J_\ell J_{\ell+2}\{-\bar{\alpha}\beta - i|\beta|^2\}e^{-2i\phi}\right\}\right]$$

$$= -\frac{\varepsilon_0 \omega}{2c}E_0^2 J_\ell^2\, \text{Im}\left[\frac{k_t^2}{2kk_z}\{\alpha\bar{\beta} - \bar{\alpha}\beta\} + \frac{k_z k_t^2}{2k^3}\{\alpha\bar{\beta} - \bar{\alpha}\beta\}\right]$$

$$= -\frac{I\omega}{2c^2}J_\ell^2\left[\frac{k_t^2}{kk_z} + \frac{k_z k_t^2}{k^3}\right]\sin 2\eta.$$

(1.44)

Finally, we calculate the pure second order transverse contribution which is fourth order in the smallness parameter.

$$C(\mathbf{T}_2 \cdot \mathbf{T}_2) = -\frac{\varepsilon_0 \omega}{2} \text{Im}\left[\hat{\mathbf{x}} \frac{k_t^2}{4k^2}\left[2\bar{\alpha} J_\ell + J_{\ell-2}\{\bar{\alpha} - i\bar{\beta}\} e^{2i\phi} + J_{\ell+2}\{\bar{\alpha} + i\bar{\beta}\} e^{-2i\phi}\right]\right.$$

$$+ \hat{\mathbf{y}} \frac{k_t^2}{4k^2}\left[2\bar{\beta} J_\ell + J_{\ell-2}\{-i\bar{\alpha} - \bar{\beta}\} e^{2i\phi} + J_{\ell+2}\{-\bar{\beta} + i\bar{\alpha}\} e^{-2i\phi}\right] E_0 \, e^{-i(k_z z + \ell\phi)}$$

$$\cdot \frac{k_t^2}{4kk_z}\left(\hat{\mathbf{x}}\left[-2\beta J_\ell + J_{\ell-2}\{i\alpha - \beta\} e^{-2i\phi} + J_{\ell+2}\{-i\alpha - \beta\} e^{2i\phi}\right]\right.$$

$$\left.\left.+ \hat{\mathbf{y}}\left[2\alpha J_\ell + J_{\ell-2}\{-i\beta - \alpha\} e^{-2i\phi} + J_{\ell+2}\{i\beta - \alpha\} e^{2i\phi}\right]\right) B_0 \, e^{i(k_z z + \ell\phi)}\right]$$

$$= -\frac{\varepsilon_0 \omega}{2c} E_0^2 \, \text{Im}\left[\frac{k_t^4}{16k^3 k_z}\left(-4\bar{\alpha}\beta J_\ell^2 + \{\bar{\alpha} - i\bar{\beta}\}\{i\alpha - \beta\} J_{\ell-2}^2 + \{\bar{\alpha} + i\bar{\beta}\}\{-i\alpha - \beta\} J_{\ell+2}^2\right.\right.$$

$$\left.\left.+ 4\alpha\bar{\beta} J_\ell^2 + \{i\bar{\alpha} - \bar{\beta}\}\{-i\beta - \alpha\} J_{\ell-2}^2 + \{-i\bar{\alpha} - \bar{\beta}\}\{i\beta - \alpha\} J_{\ell+2}^2\right)\right]$$

$$= -\frac{\varepsilon_0 \omega}{2c} E_0^2 \, \text{Im}\left[\frac{k_t^4}{16k^3 k_z}\left(4(\alpha\bar{\beta} - \bar{\alpha}\beta) J_\ell^2 + \{i|\alpha|^2 + \alpha\bar{\beta} - \bar{\alpha}\beta + i|\beta|^2\} J_{\ell-2}^2\right.\right.$$

$$\left.\left.+ \{-i|\alpha|^2 + \alpha\bar{\beta} - \bar{\alpha}\beta - i|\beta|^2\} J_{\ell+2}^2 + \{-\bar{\alpha}\beta + i|\beta|^2 + i|\alpha|^2 + \alpha\bar{\beta}\} J_{\ell-2}^2 + \{-i|\beta|^2 - \bar{\alpha}\beta + \alpha\bar{\beta} - i|\alpha|^2\} J_{\ell+2}^2\right)\right]$$

$$= -\frac{I\omega}{c^2} \frac{k_t^4}{8k^3 k_z} \text{Im}\left[\left(2(\alpha\bar{\beta} - \bar{\alpha}\beta) J_\ell^2 + \{i|\alpha|^2 + \alpha\bar{\beta} - \bar{\alpha}\beta + i|\beta|^2\} J_{\ell-2}^2 + \{-i|\alpha|^2 + \alpha\bar{\beta} - \bar{\alpha}\beta - i|\beta|^2\} J_{\ell+2}^2\right)\right]$$

$$= -\frac{I\omega}{c^2} \frac{k_t^4}{8k^3 k_z}\left[\left(2(\sin 2\eta) J_\ell^2 + \{\sin 2\eta + 1\} J_{\ell-2}^2 + \{\sin 2\eta - 1\} J_{\ell+2}^2\right)\right].$$

(1.45)

Thus adding up (1.42), (1.43), (1.44), and (1.45) leads to the total optical chirality density:

$$C = -\frac{I\omega}{c^2}\left[\frac{k_z}{k} J_\ell^2 \sin 2\eta + \frac{k_t^2}{4kk_z}\left(\left(1 + \frac{k_z^2}{k^2}\right) 2 J_\ell^2 \sin 2\eta + \{\sin 2\eta + 1\} J_{\ell-1}^2 + \{\sin 2\eta - 1\} J_{\ell+1}^2\right)\right.$$

$$\left.+ \frac{k_t^4}{8k^3 k_z}\left[\left(2 J_\ell^2 \sin 2\eta + \{\sin 2\eta + 1\} J_{\ell-2}^2 + \{\sin 2\eta - 1\} J_{\ell+2}^2\right)\right]\right]. \quad (1.46)$$

(1.46) is a quadratic field quantity and therefore can be written as a sum of polarized and unpolarized part. We may introduce the degree of polarization $0 \leq P \leq 1$ such that

$$C = -\frac{I\omega}{c^2}\left[\frac{k_z}{k} PJ_\ell^2 \sin 2\eta + \frac{k_t^2}{4kk_z}\left(\left(1 + \frac{k_z^2}{k^2}\right) 2 P J_\ell^2 \sin 2\eta + \{P\sin 2\eta + 1\} J_{\ell-1}^2 + \{P\sin 2\eta - 1\} J_{\ell+1}^2\right)\right.$$

$$\left.+ \frac{k_t^4}{8k^3 k_z}\left[\left(2 P J_\ell^2 \sin 2\eta + \{P\sin 2\eta + 1\} J_{\ell-2}^2 + \{P\sin 2\eta - 1\} J_{\ell+2}^2\right)\right]\right]. \quad (1.47)$$

We note that $P\sin 2\eta$ is essentially the third stokes parameter for partially polarized light. The other terms which do not depend on the ellipticity are polarization-independent, and are present whether the input light is linearly, elliptically, circularly, or even unpolarized as we highlight in the main manuscript.

For example, it is trivial to show a completely 2D polarized beam $P=1$ gives the identical result as (1.46), while a 2D unpolarized beam $P=0$ has the following chirality density

$$C = -\frac{I\omega}{c^2}\left[\frac{k_t^2}{4kk_z}\left(J_{\ell-1}^2 - J_{\ell+1}^2\right) + \frac{k_t^4}{8k^3k_z}\left[J_{\ell-2}^2 - J_{\ell+2}^2\right]\right]. \tag{1.48}$$

Which does obviously not depend on the ellipticity either as it shouldn't.

## Additional chirality density distributions for purely polarized light *P*=1

Here we provide additional plots of the spatial distribution for purely polarized Bessel beams relevant to Section 2.1 of the main manuscript. In particular Figs S1 and S2 show the spatial distribution of the optical chirality density Equation (6) from the main manuscript for cases of $|\ell|=2$, i.e. showing the effects of increasing the OAM of the input beam.

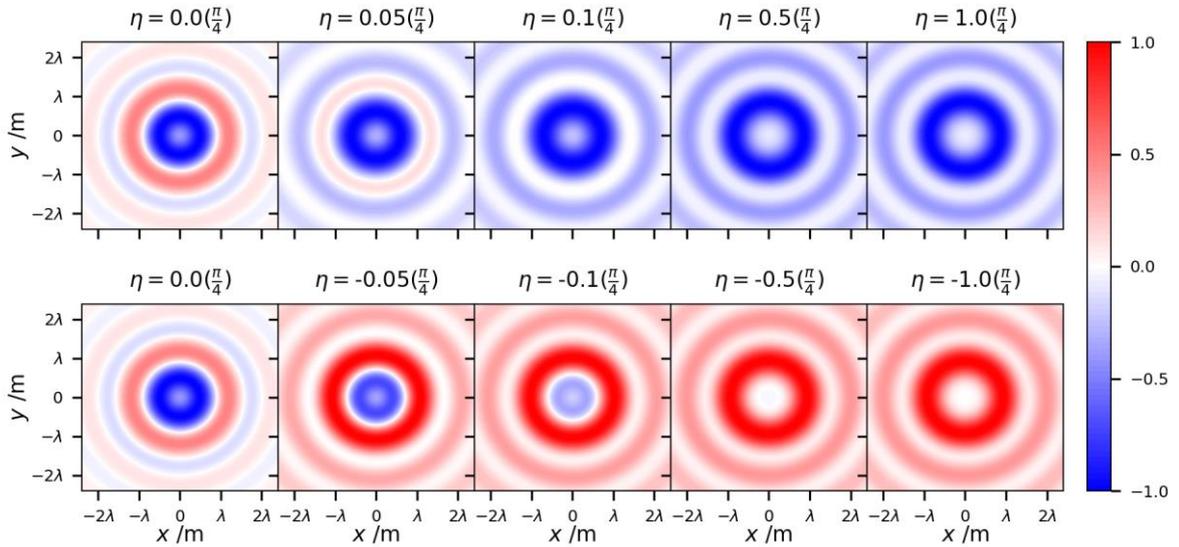

**Figure S1:** Evolution of the optical chirality density of an $\ell=2$ Bessel beam in the focal plane with varying 2D-polarization ellipticity $\eta$. The input polarization progresses from linearly polarized $\eta=0$ through varying degrees of ellipticity $-\pi/4 < \eta < \pi/4$ until it reaches pure circularly polarization $\eta=\pm\pi/4$. The top row cycles through right-handed polarization; the bottom row left-handed polarization. In all plots $k_t/k_z = 0.6315$ and each plot is normalized individually.

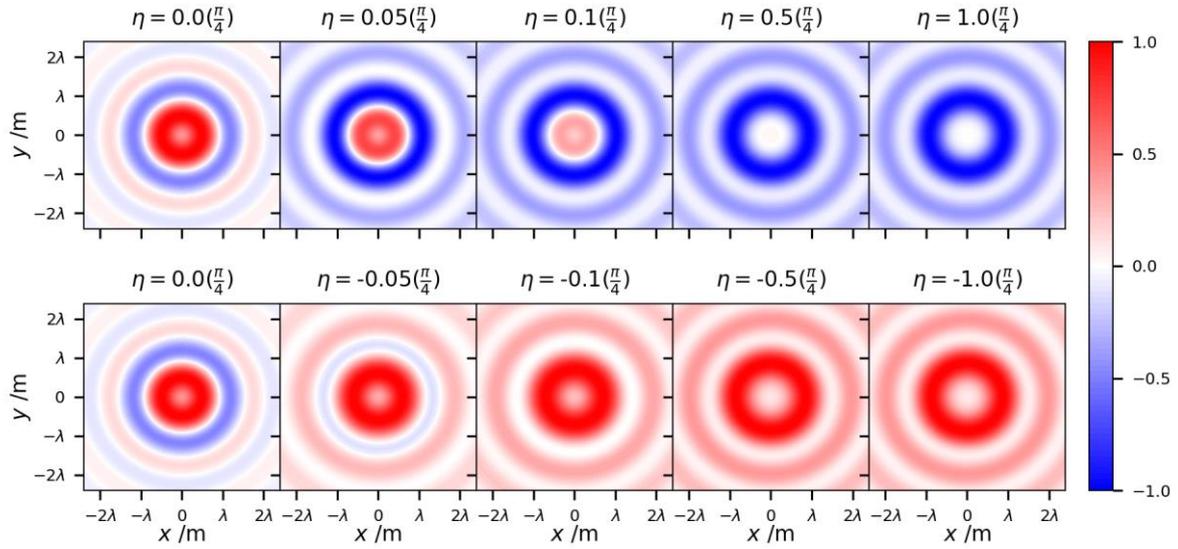

**Figure S2:** Evolution of the optical chirality density of an $\ell = -2$ Bessel beam in the focal plane with varying 2D-polarization ellipticity $\eta$. The input polarization progresses from linearly polarized $\eta = 0$ through varying degrees of ellipticity $-\pi/4 < \eta < \pi/4$ until it reaches pure circularly polarization $\eta = \pm\pi/4$. The top row cycles through right-handed polarization; the bottom row left-handed polarization. In all plots $k_t/k_z = 0.6315$ and each plot is normalized individually.

## Additional chirality density distributions for partially polarized light

Figure S3 correspond to Figs 5-7 in the main manuscript with $\eta = \pm\pi/4$. The result for $\eta = 0$ is trivial and just repeats the 2D linear/2D unpolarized spatial distribution across all values of $P$.

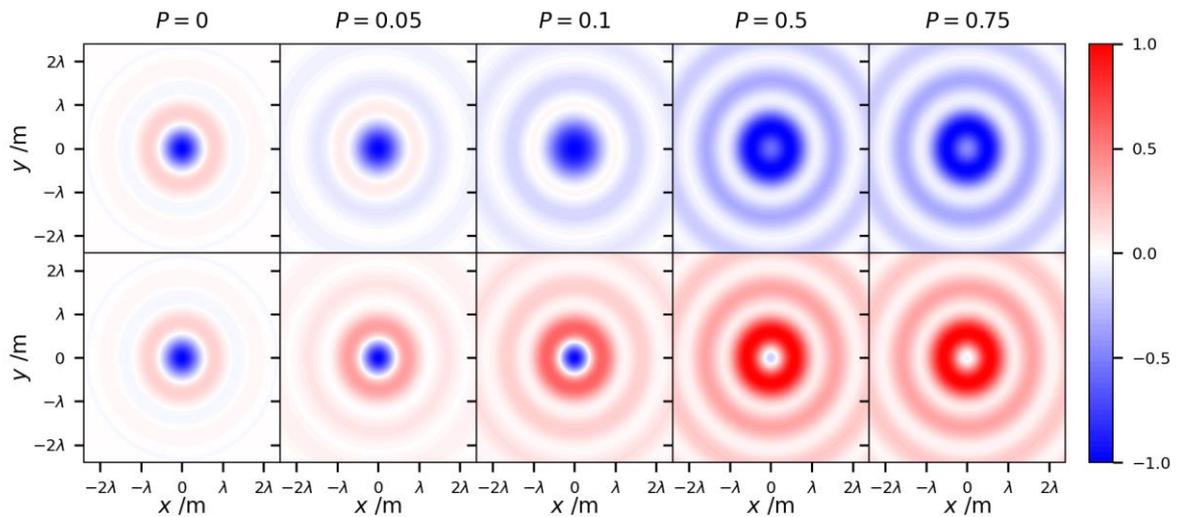

**Figure S3:** Evolution of the optical chirality density of an $\ell = 1$ Bessel beam in the focal plane with varying degree of polarization $P$ of an elliptically polarized beam $\eta = \pm\pi/4$. The top row cycles through right-handed polarization; the bottom row left-handed polarization. In all plots $k_t/k_z = 0.6315$ and each plot is normalized individually.

The Figures S4-S7 correspond to Figure 5-7 from the main manuscript and Figure S3 but for $\ell = -1$.

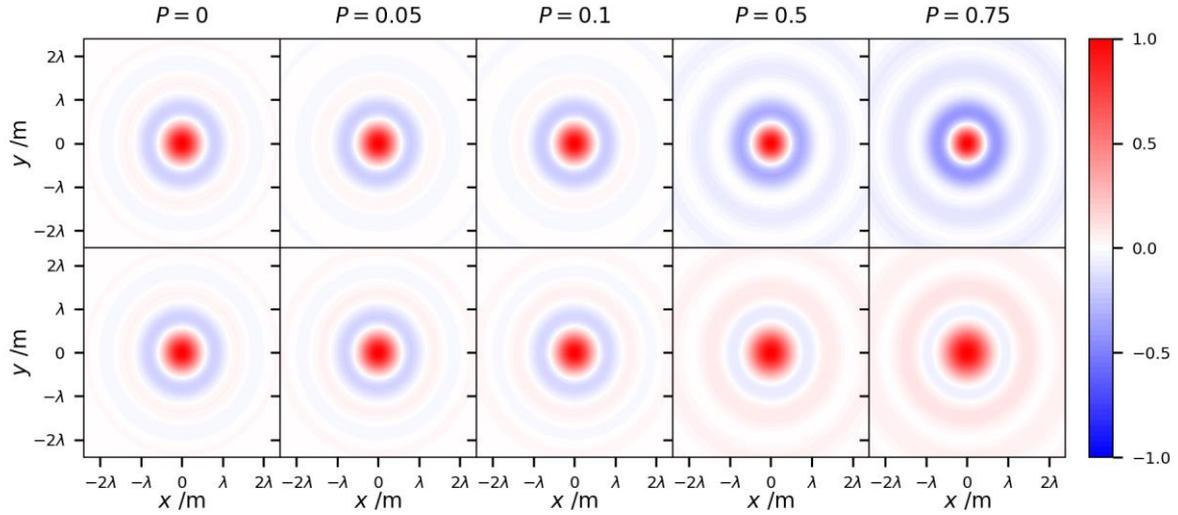

**Figure S4:** Evolution of the optical chirality density of an $\ell = -1$ Bessel beam in the focal plane with varying degree of polarization *P* of an elliptically polarized beam $\eta = \pm\pi/80$. The top row cycles through right-handed polarization; the bottom row left-handed polarization. In all plots $k_t / k_z = 0.6315$ and each plot is normalized individually.

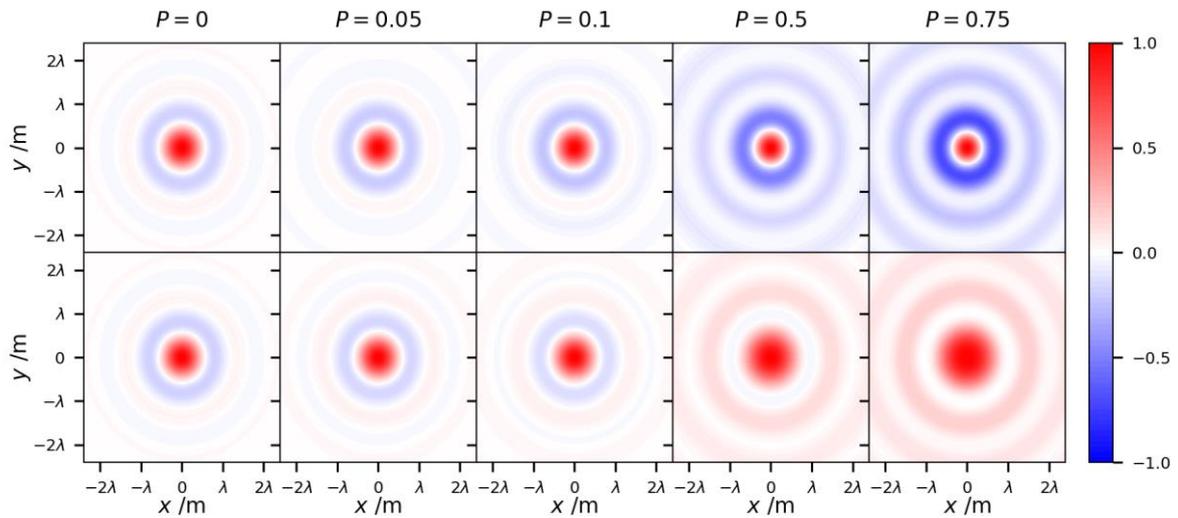

**Figure S5:** Evolution of the optical chirality density of an $\ell = -1$ Bessel beam in the focal plane with varying degree of polarization *P* of an elliptically polarized beam $\eta = \pm\pi/40$. The top row cycles through right-handed polarization; the bottom row left-handed polarization. In all plots $k_t / k_z = 0.6315$ and each plot is normalized individually.

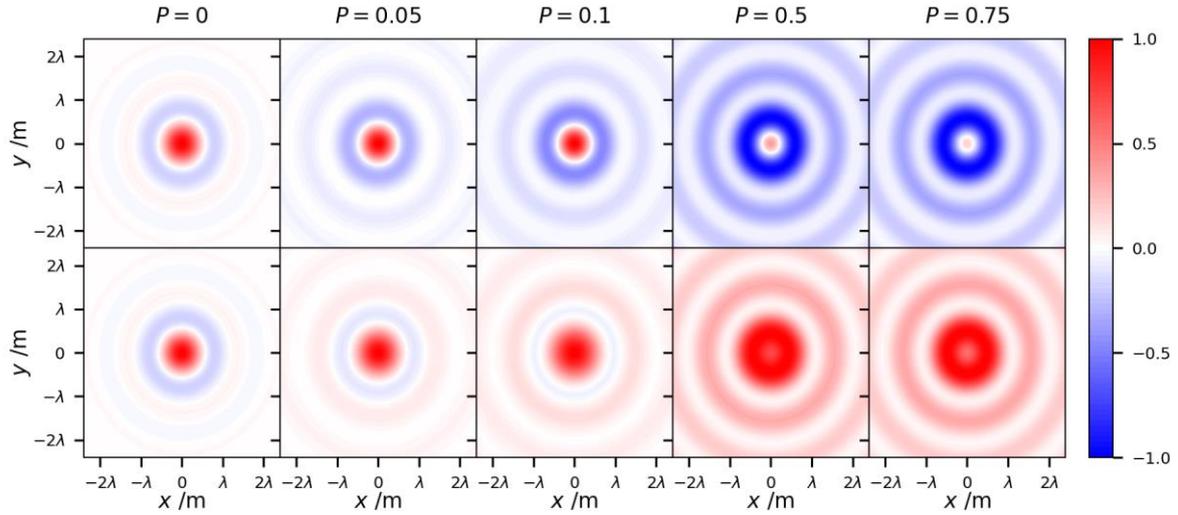

**Figure S6:** Evolution of the optical chirality density of an $\ell = -1$ Bessel beam in the focal plane with varying degree of polarization *P* of an elliptically polarized beam $\eta = \pm\pi/8$. The top row cycles through right-handed polarization; the bottom row left-handed polarization. In all plots $k_t/k_z = 0.6315$ and each plot is normalized individually.

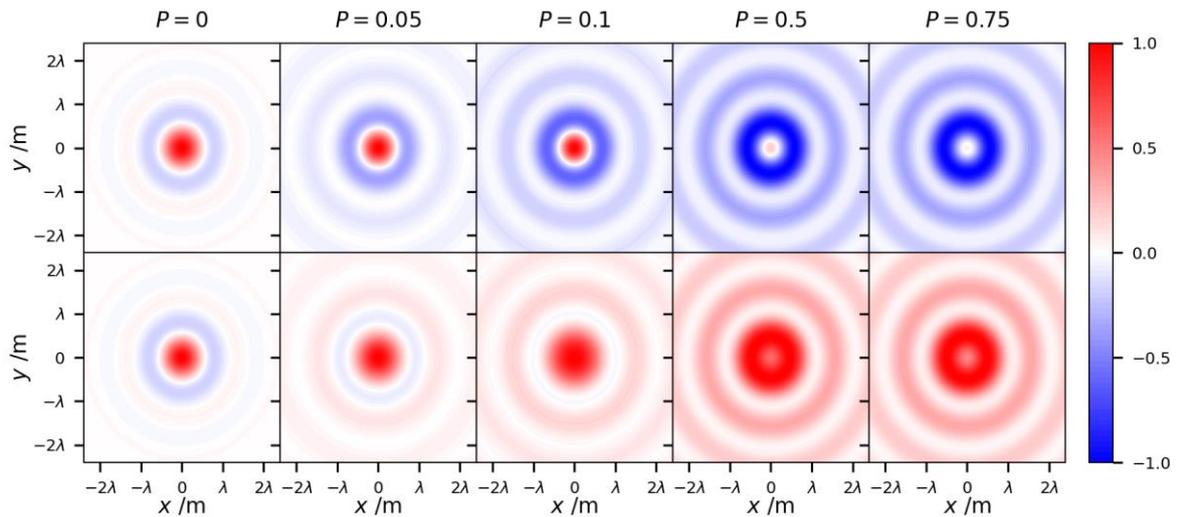

**Figure S7:** Evolution of the optical chirality density of an $\ell = -1$ Bessel beam in the focal plane with varying degree of polarization *P* of an elliptically polarized beam $\eta = \pm\pi/4$. The top row cycles through right-handed polarization; the bottom row left-handed polarization. In all plots $k_t/k_z = 0.6315$ and each plot is normalized individually.

Finally, the plots below are some examples of how increasing the magnitude of $\ell$ along with varying the degree of polarization influences the optical chirality density.

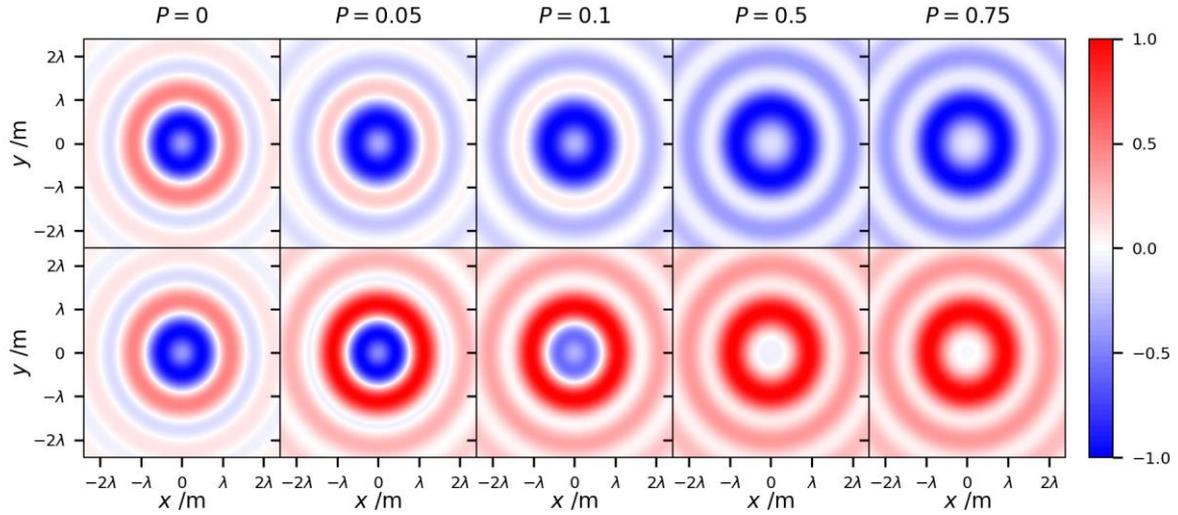

**Figure S8:** Evolution of the optical chirality density of an $\ell = 2$ Bessel beam in the focal plane with varying degree of polarization *P* of an elliptically polarized beam $\eta = \pm \pi / 4$. The top row cycles through right-handed polarization; the bottom row left-handed polarization. In all plots $k_t / k_z = 0.6315$ and each plot is normalized individually.

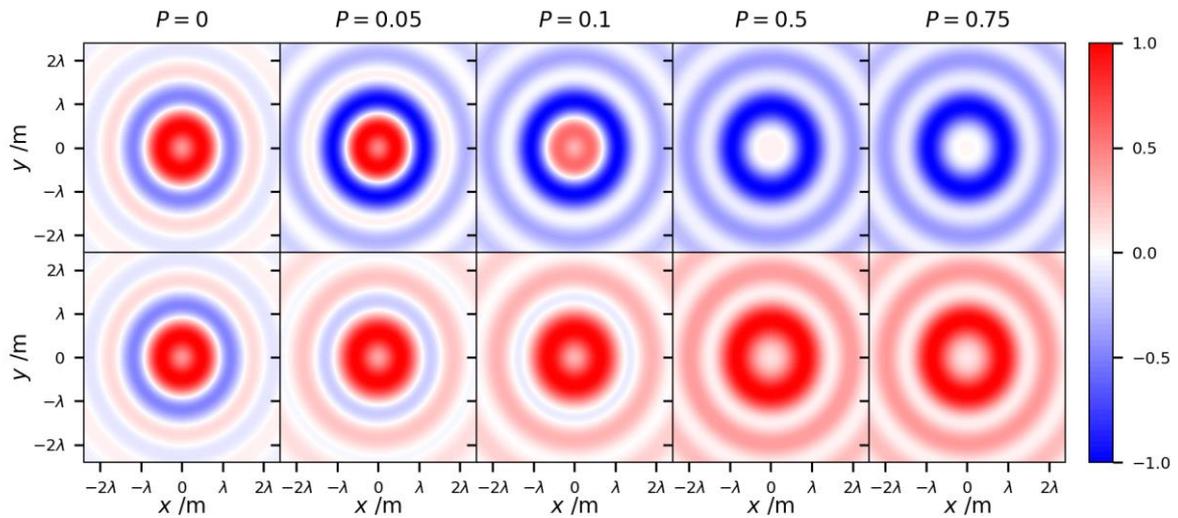

**Figure S9:** Evolution of the optical chirality density of an $\ell = -2$ Bessel beam in the focal plane with varying degree of polarization *P* of an elliptically polarized beam $\eta = \pm \pi / 4$. The top row cycles through right-handed polarization; the bottom row left-handed polarization. In all plots $k_t / k_z = 0.6315$ and each plot is normalized individually.

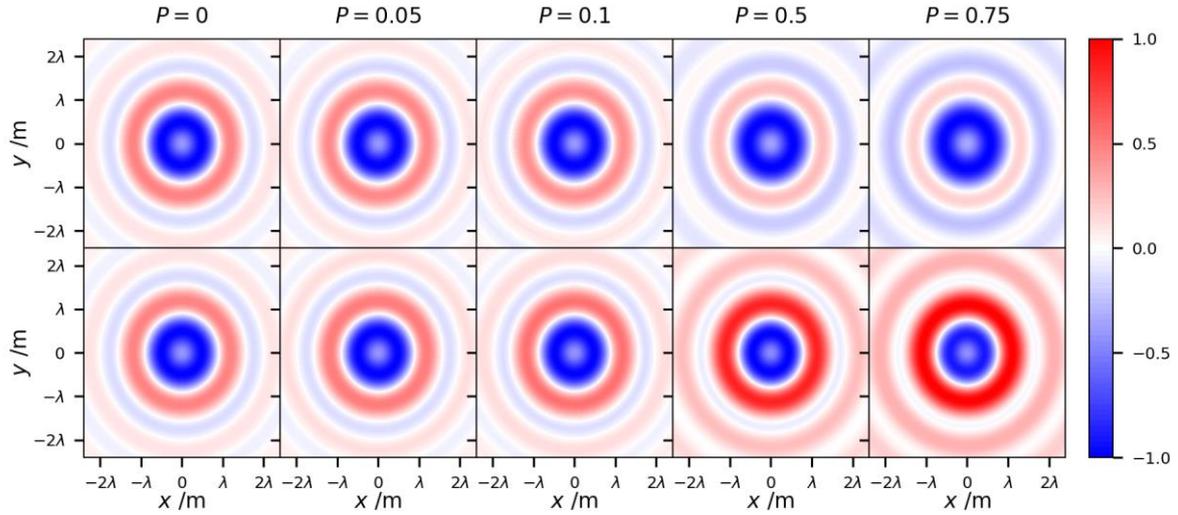

**Figure S10:** Evolution of the optical chirality density of an $\ell = 2$ Bessel beam in the focal plane with varying degree of polarization $P$ of an elliptically polarized beam $\eta = \pm\pi/80$. The top row cycles through right-handed polarization; the bottom row left-handed polarization. In all plots $k_t/k_z = 0.6315$ and each plot is normalized individually.

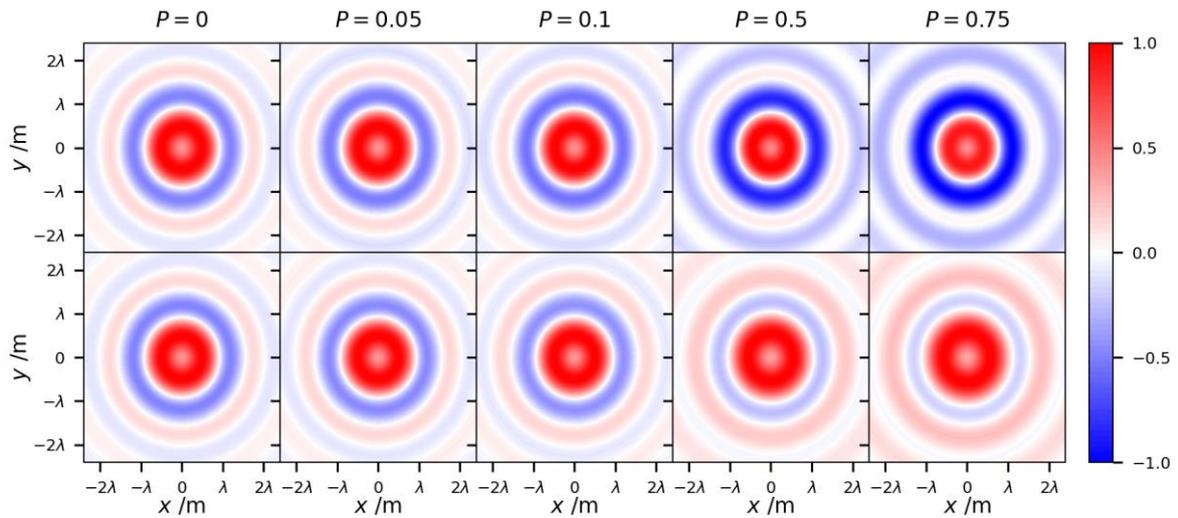

**Figure S11:** Evolution of the optical chirality density of an $\ell = -2$ Bessel beam in the focal plane with varying degree of polarization $P$ of an elliptically polarized beam $\eta = \pm\pi/80$. The top row cycles through right-handed polarization; the bottom row left-handed polarization. In all plots $k_t/k_z = 0.6315$ and each plot is normalized individually.